\documentclass[floats,floatfix,showpacs,amssymb,prd,twocolumn,superscriptaddress,nofootinbib]{revtex4-1}

\usepackage{amssymb,amsmath,verbatim,mathtools,needspace,enumitem,etoolbox,graphicx,physics,microtype}

\usepackage[dvipsnames, usenames]{xcolor}
\usepackage[unicode, colorlinks=true, linkcolor=linkcolor, citecolor=linkcolor, filecolor=linkcolor,urlcolor=linkcolor, pdfusetitle]{hyperref}
\usepackage[all]{hypcap}
\usepackage[T1]{fontenc}
\usepackage[utf8]{inputenc}

\usepackage[utf8]{inputenc}
\usepackage{graphicx,amssymb,amsmath,amsthm,amsfonts,epsfig,times,natbib}

\usepackage{epstopdf}
\usepackage{tensor}
\usepackage{amsbsy}
\usepackage{bm,url}
\usepackage{float}
\usepackage{dcolumn}
\usepackage{latexsym}
\usepackage{rotating}
\usepackage{longtable}
\usepackage{enumerate}
\usepackage{booktabs}
\usepackage[utf8]{inputenc}
\usepackage{gensymb}

\usepackage{bm}
\usepackage[text={7in,9in},centering]{geometry}
\usepackage{aas_macros}

\newcolumntype{L}{>{$}l<{$}}
\newcolumntype{R}{>{$}r<{$}}
\newcolumntype{C}{>{$}c<{$}}

\definecolor{oxfordblue}{rgb}{0.0, 0.13, 0.28}
\definecolor{burgundy}{rgb}{0.5, 0.0, 0.13}
\definecolor{darkolivegreen}{rgb}{0.33, 0.42, 0.18}
\definecolor{darkblue}{rgb}{0,0,0.5}
\definecolor{richcarmine}{rgb}{0.84, 0.0, 0.25}
\definecolor{darkblue}{rgb}{0,0,0.5}
\definecolor{venetianred}{rgb}{0.78, 0.03, 0.08}
\definecolor{skobeloff}{rgb}{0.0, 0.48, 0.45}
\hypersetup{colorlinks=true, citecolor=darkblue, linkcolor=darkblue,
urlcolor = darkblue}

\newcommand{\jhu}{\affiliation{Department of Physics and Astronomy, Johns Hopkins University, 3400 N. Charles
Street, Baltimore, MD 21218, USA}}

\newcommand{\lm}[1]{\ensuremath{{#1}_{\ell m}}}

\def\nn{\nonumber}

\newcommand{\ben}{\begin{enumerate}}
\newcommand{\een}{\end{enumerate}}

\def\cov{{\rm cov}}
\def\lm{{\ell m}}

\newcommand{\jacob}[2]{{\frac{\partial #1}{\partial #2}}}

\def\be{\begin{equation}}
\def\ee{\end{equation}}
\def\bea{\begin{eqnarray}}
\def\eea{\end{eqnarray}}
\def\nn{\nonumber}
\newcommand{\beq}{\begin{eqnarray}}
\newcommand{\eeq}{\end{eqnarray}} 
\newcommand{\ba}{\begin{align}}
\newcommand{\ea}{\end{align}}

\def\nn{\nonumber}
\def\be{\begin{equation}}
\def\ee{\end{equation}}
\def\beq{\begin{eqnarray}}
\def\eeq{\end{eqnarray}}

\def\f{\frac}
\begin{document}

\title{LISA parameter estimation and source localization with higher harmonics of the ringdown}

\author{Vishal Baibhav}
\email{vbaibha1@jhu.edu}
\jhu

\author{Emanuele Berti}
\email{berti@jhu.edu}
\jhu

\author{Vitor Cardoso}
\email{vitor.cardoso@ist.utl.pt}
\affiliation{CENTRA, Departamento de F\'{\i}sica, Instituto Superior T\'ecnico -- IST, Universidade de Lisboa -- UL,
Avenida Rovisco Pais 1, 1049 Lisboa, Portugal}


\date{\today}

\begin{abstract}
LISA can detect higher harmonics of the ringdown gravitational-wave signal from massive black-hole binary mergers with large signal-to-noise ratio. The most massive black-hole binaries are more likely to have electromagnetic counterparts, and the inspiral will contribute little to their signal-to-noise ratio. Here we address the following question: can we extract the binary parameters and localize the source using LISA observations of the ringdown only? Modulations of the amplitude and phase due to LISA’s motion around the Sun can be used to disentangle the source location and orientation when we detect the long-lived inspiral signal, but they can not be used for ringdown-dominated signals, which are very short-lived.  We show that (i) we can still measure the mass ratio and inclination of high-mass binaries by carefully combining multiple ringdown harmonics, and (ii) we can constrain the sky location and luminosity distance by relying on the relative amplitudes and phases of various harmonics, as measured in different LISA channels.
\end{abstract}

\maketitle
\section{Introduction}

Gravitational waves are predominantly quadrupolar. For the black hole (BH) binaries detected by LIGO and Virgo, the fraction of energy radiated in subdominant multipoles increases with the mass ratio $q$~\cite{Buonanno:2006ui,Berti:2007fi} (we define $q\equiv m_1/m_2\geq 1$, where $m_1$ is the mass of the primary and $m_2$ is the mass of the secondary). For BH binaries of total mass $M=m_1+m_2$, gravitational-wave frequencies scale like $1/M$.
Simple WKB arguments~\cite{Press:1971wr} suggest that the quasinormal mode frequencies of the remnant are roughly proportional to the harmonic index $\ell$ (see e.g.~\cite{Kokkotas:1999bd,Nollert:1999ji,Berti:2009kk} for reviews).
Since higher multipoles corresponds to higher harmonics of the ringdown signal, which radiate at higher frequencies, high-$\ell$ modes become more important for high-mass binaries.

Interest in higher harmonics is growing as the sensitivity of interferometric detectors improves~\cite{OShaughnessy:2017tak,Kumar:2018hml,Cotesta:2018fcv,Mehta:2019wxm,Breschi:2019wki,Shaik:2019dym}.
This is because (if detectable) subdominant multipoles and higher harmonics of the radiation add structure to the gravitational waveforms. Different harmonics have different dependence on inclination, mass ratio and spins, so their observation can break some of the degeneracies that currently haunt the parameter estimation.

\begin{figure*}[t]
\begin{tabular}{c}
  \includegraphics[width=0.98\textwidth]{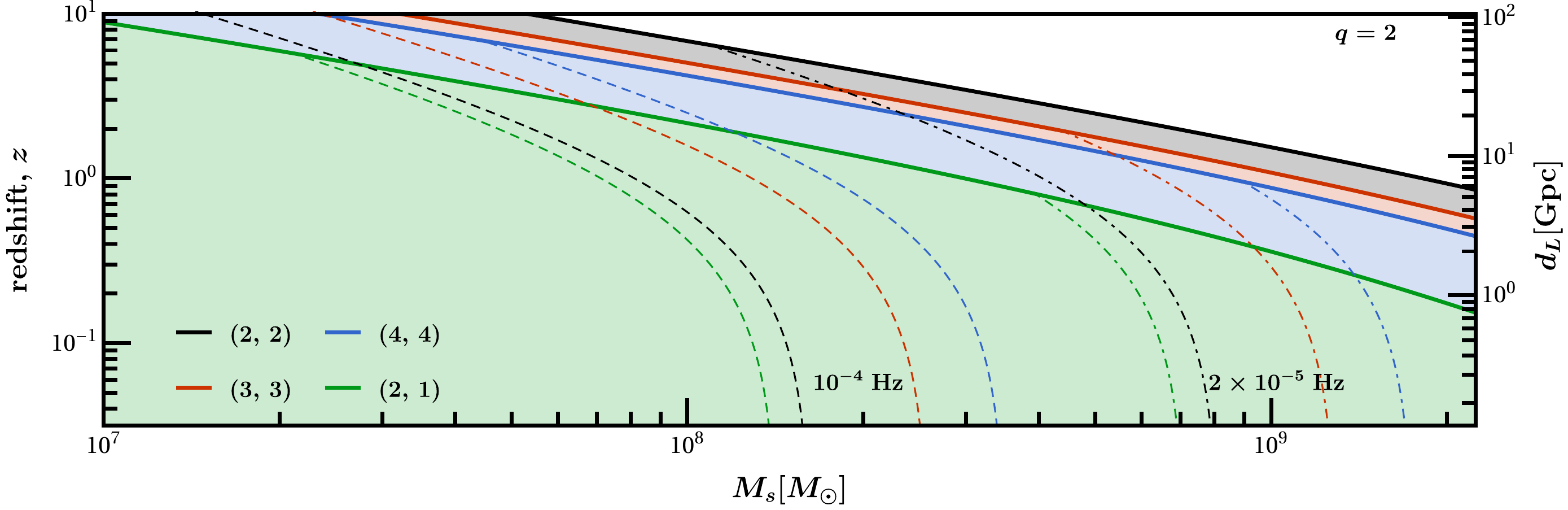}\\
  \includegraphics[width=0.98\textwidth]{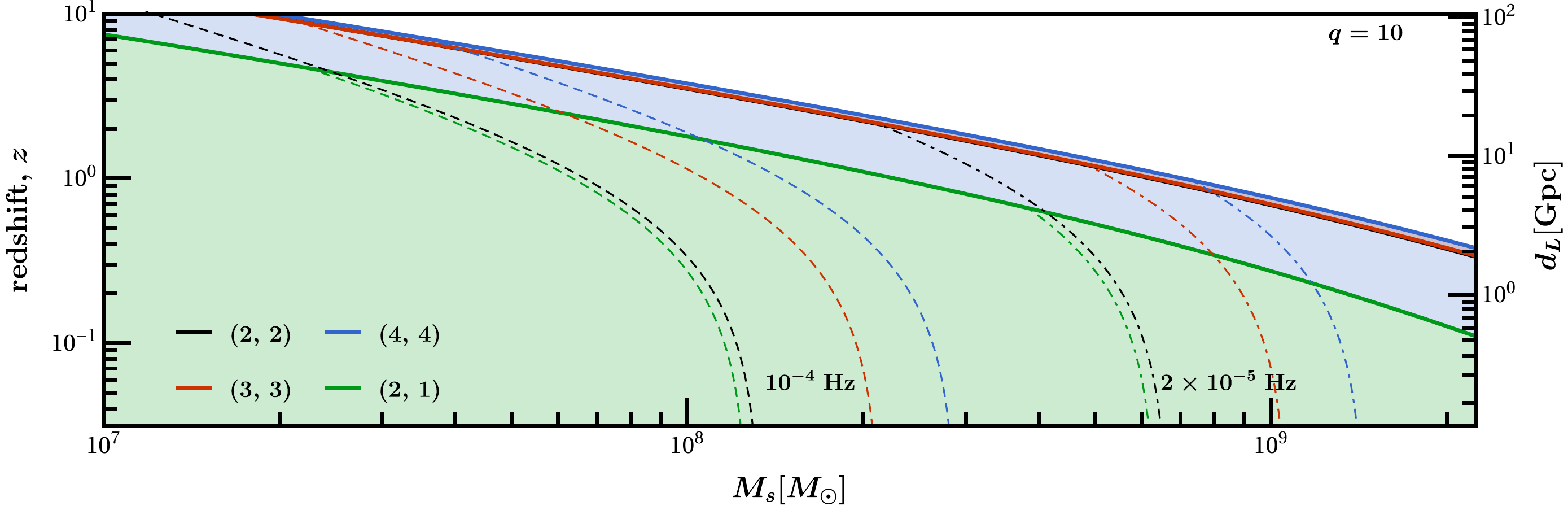}
  \end{tabular}
 \caption{Solid lines indicate ringdown horizons for $(2,\,2)$, $(3,\,3)$, $(2,\,1)$, $(4,\,4)$ modes for a binary with $q=2$ (top) and  $q=10$ (bottom). Dashed and dash-dotted lines correspond to a low-frequency cutoff $f_{\rm cut}=10^{-4}$~Hz and $f_{\rm cut}=2\times10^{-5}$~Hz, respectively.
 }
\label{fig:CRTS}
\end{figure*}

One example is the distance-inclination degeneracy. Different multipoles correspond to different spherical harmonic indices and to a different angular dependence (and hence inclination dependence) of the radiation. Therefore higher multipoles allow us to distinguish between different binary orientations, and this can also lead to improvements in distance measurements. Degeneracy breaking can also occur because the excitation of each higher multipole depends in a characteristic way on the mass ratio $q$ and on the spins~\cite{Barausse:2009xi,Pan:2010hz,Kamaretsos:2011um,Kamaretsos:2012bs,London:2014cma,Baibhav:2017jhs,Baibhav:2018rfk}. This can break the degeneracy between the mass ratio $q$ and the so-called ``effective spin'' parameter $\chi_{\rm eff}$. For example, it was recently shown that higher harmonics allow us to better determine the mass ratio of the most massive BH binary detected to date (GW170729)~\cite{Chatziioannou:2019dsz}, and this can also lead to improved effective spin estimates.  Higher-order modes can also break the degeneracy between polarization and coalescence phase~\cite{Payne:2019wmy}.

In this paper we will focus on the information carried by higher multipoles of the ringdown, as they may be detectable by the space-based interferometer LISA~\cite{Audley:2017drz}. Several works have studied how LISA detectability and parameter estimation are affected by higher harmonics {\em of the inspiral}, finding that they can improve LISA's angular resolution and (consequently) luminosity distance estimates by a factor $\sim 10^2$, especially for heavier binaries with $M\gtrsim 10^7 M_\odot$~\cite{Arun:2007hu,Trias:2007fp,Arun:2007qv,Porter:2008kn}.

Ringdown is expected to be dominant over the inspiral for binaries with mass $M\gtrsim 10^6 M_\odot$~\cite{Berti:2005ys,Flanagan:1997sx,Rhook:2005pt}. Higher harmonics of the signal usually have low amplitudes during the inspiral, and become dominant only during merger and ringdown (see e.g.~\cite {Bustillo:2015qty}). In general, higher harmonics are more important in the ringdown stage: during the inspiral the higher harmonics are always subdominant relative to the inspiral of the $(2,\,2)$ mode, while harmonics with $\ell=m>2$ stand out in the frequency domain during the ringdown, because they have larger frequencies (and hence are not  ``buried'' under the $(2,\,2)$ component of the signal).

Since higher multipoles typically correspond to higher frequencies and $f\sim 1/M$, when $M$ is large enough the dominant mode will fall out of the sensitivity band of LISA and become undetectable: higher harmonics could be our only means to observe otherwise undetectable high-mass sources. For systems with mass $M\gtrsim 10^6 M_\odot$, high-frequency harmonics can lie closer to the noise ``bucket'' of LISA than the fundamental (low-frequency) modes, and therefore they can have relatively large SNR. This is particularly important for large-$q$ mergers, because then higher modes can have relatively large amplitudes relative to the $(2,\,2)$ mode~\cite{Kamaretsos:2011um,Baibhav:2017jhs,Cotesta:2018fcv}. In fact, the SNR in higher harmonics  for massive binaries with large $q$ is comparable to (or greater than) the $(2,\,2)$ mode SNR. 

It is generally believed that it will be hard to control LISA's noise below a low-frequency cut-off $f_{\rm cut}\sim 10^{-4}\;{\rm Hz}$, or possibly $f_{\rm cut}\sim 2\times 10^{-5}\;{\rm Hz}$. A low-frequency cutoff implies that there is a maximum {\em redshifted} mass $M^{\rm cut}_\lm$ beyond which the $(\ell,\,m)$ mode goes out of band. This maximum mass can be written as
\be\label{eq:Mcut}
M^{\rm cut}_{\ell m} = \mu^\lm_{8} M_{\odot} \f{10^{-4}\,\rm Hz}{f_{\rm cut}}\; \f{\hat{\omega}_{\ell m}}{\hat{\omega}^{q=1}_{\ell m}}\,.
\ee
Here $\hat{\omega}_\lm$ denotes dimensionless QNM frequencies scaled by their maximum value $\hat{\omega}^{q=1}_{\ell m}$, which for nonspinning BH binary mergers corresponds to $q=1\,(a=0.686)$. As shown in~\cite{Berti:2009kk}, these frequencies are well fitted by an expression of the form 
\be\label{eq:fq1}
\hat{\omega}_{\ell m}= f^\lm_1+f^\lm_2 (1-a)^{f^\lm_3}\,.
\ee
For mergers of nonspinning BHs, the remnant spin $a$ is a function of mass ratio $q$ only. It can be approximated as~\cite{Barausse:2009uz}
\be
a(q)=\eta  \left(2\sqrt{3}-3.5171\,\eta + 2.5763\,\eta ^2\right)\,,
\ee 
where $\eta=q/(1+q)^2$ is the symmetric mass ratio. In Table~\ref{tab:coeff} we list $\mu^\lm_{8}$, $\hat{\omega}^{q=1}_{\ell m}$, $f^\lm_1$, $f^\lm_2$ and $f^\lm_3$ for the dominant modes. 

\begin{table}
  \caption{Fitting coefficients for Eqs.~(\ref{eq:Mcut}) and (\ref{eq:fq1}).}
\label{tab:coeff}
\setlength\tabcolsep{6 pt}
\begin{tabular}{LCCCCC}
\hline
{(\ell,\,m)} & \mu^\lm_8 & \hat{\omega}^{q=1}_{\ell m} & f^\lm_1 & f^\lm_2 & f^\lm_3 \\
\hline
{(2,\,2)} & 1.71\times 10^8 & 0.529 & 1.525 & -1.157 & 0.129 \\
 {(3,\,3)} & 2.71\times 10^8 & 0.839 & 1.896 & -1.304 & 0.182 \\
 {(2,\,1)} & 1.47\times 10^8 & 0.456 & 0.6 & -0.234 & 0.418 \\
 {(4,\,4)} & 3.68\times 10^8 & 1.139 & 2.3 & -1.506 & 0.224 \\
\hline
\end{tabular}
\end{table}

The importance of the low-frequency cut-off can be appreciated by looking at Fig.~\ref{fig:CRTS}, where we consider nonspinning binary mergers with $q=2$ (top panel) and $q=10$ (bottom panel). Low-frequency sensitivity is crucial to observe ringdown from the most massive BH mergers, so we also plot ringdown horizons obtained by truncating the LISA noise power spectral density at $f_{\rm cut}=10^{-4}$~Hz (dashed lines) and $f_{\rm cut}=2\times10^{-5}$~Hz (dash-dotted lines). LISA Pathfinder exceeded the LISA requirements at frequencies as low as $2\times10^{-5}$ Hz~\cite{Armano:2018kix}. If the LISA constellation noise can be trusted at these same frequencies,  the mass reach of the instrument would extend up to $\sim10^9 M_\odot$, where the inspiral is not visible and most of the SNR will come from merger and ringdown.

\subsection{Plan of the paper}
In this work we study LISA parameter estimation using only the ringdown. The various sections address the measurement of different parameters, as follows:

\noindent
{\bf Remnant mass and spin.} The spin and (redshifted) mass of the remnant can be found from measurements of the quasinormal mode frequencies. In Sec.~\ref{sec:remnant} we study how accurately LISA can measure the remnant mass and spin, and how higher harmonics can improve these measurements.

\noindent
{\bf Mass ratio and inclination.} The relative excitation of higher multipoles depends on the binary mass ratio $q$ and inclination angle $\iota$. In Sec.~\ref{sec:qiota} we use estimates of the relative amplitudes of different $\ell=m$ modes to measure $q$ and $\iota$.

\noindent
{\bf Source location and luminosity distance.} LISA inspiral sources are long-lived, and LISA's motion around the Sun modulates the amplitude and phase of the signal, which in turn can be used to disentangle the source location and orientation. On the contrary, the ringdown is very short-lived, and hence we cannot use the modulation of the antenna pattern for localization. Furthermore, the angular dependence of different modes with $\ell=m$ depends only on $\iota$, so we must rely on modes with $\ell\neq m$ to infer more information on the source location. In Secs.~\ref{sec:skyloc} and ~\ref{sec:lumdistance} we show that we can constrain the sky location and luminosity distance by relying on the relative amplitudes and phases of the $(2,\,2)$ and $(2,\,1)$ modes, as measured in different LISA channels.

In Sec.~\ref{sec:qIotaSkyDep} we present a preliminary exploration of the dependence of the errors on mass ratio, inclination, and sky-location.

In Sec.~\ref{sec:conclusions} we summarize our results and discuss possible directions for future work.

In most of this paper we ignore the motion of LISA, because ringdown signals are typically much shorter than LISA's observation time and orbital period. This assumption is justified in Appendix~\ref{sec:evolutionLocalization}, where we study the effect of first-order corrections to this approximation. We show that these corrections are negligible even for binaries with $M>10^8 M_\odot$, when the ringdown can last for hours. Finally, in Appendix~\ref{sec:EM} we show that parameter estimation could improve dramatically for sources that can be associated with an electromagnetic counterpart.

\begin{figure*}[t]
\begin{tabular}{cc}
  \includegraphics[width=0.45\textwidth]{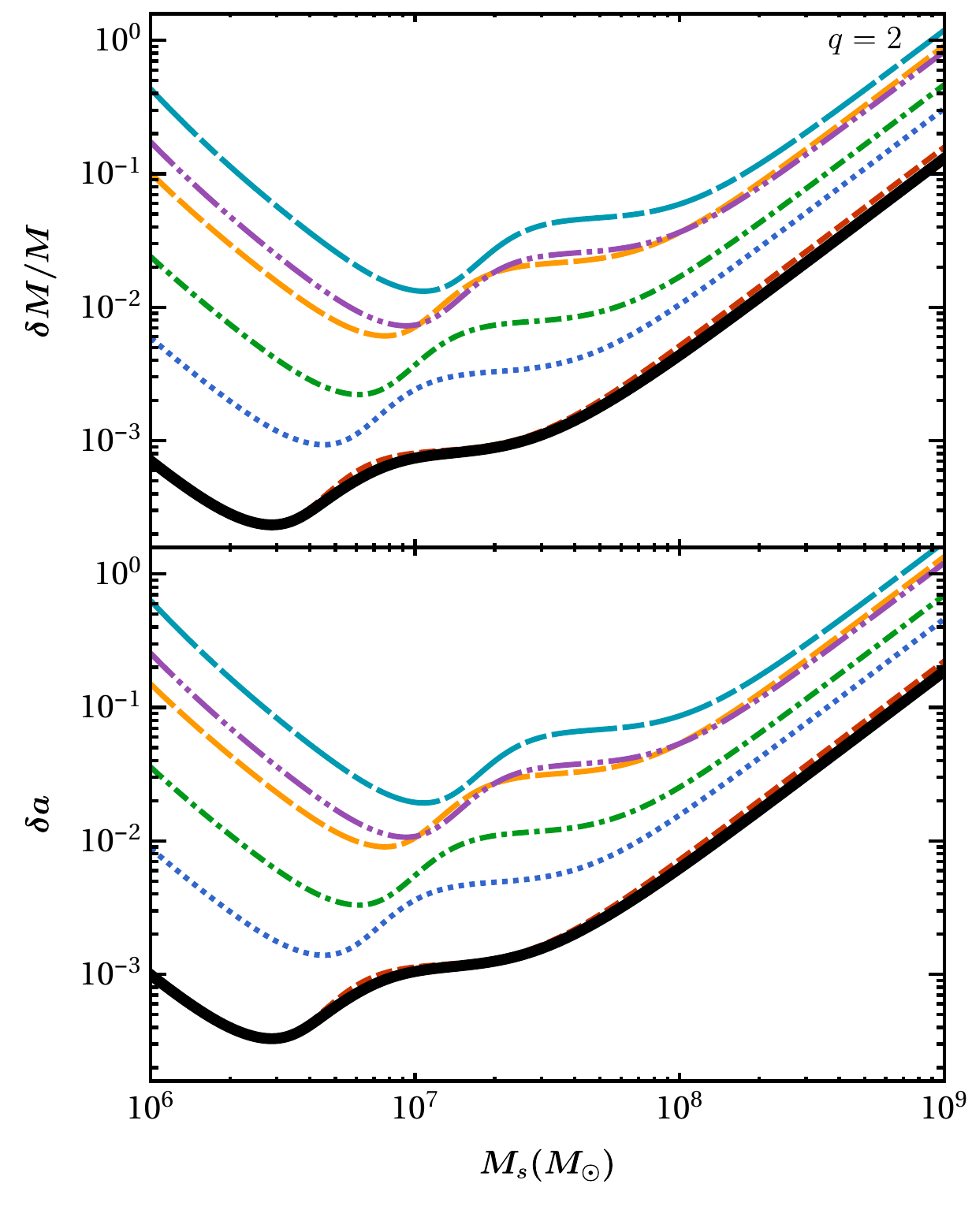}&  
  \includegraphics[width=0.45\textwidth]{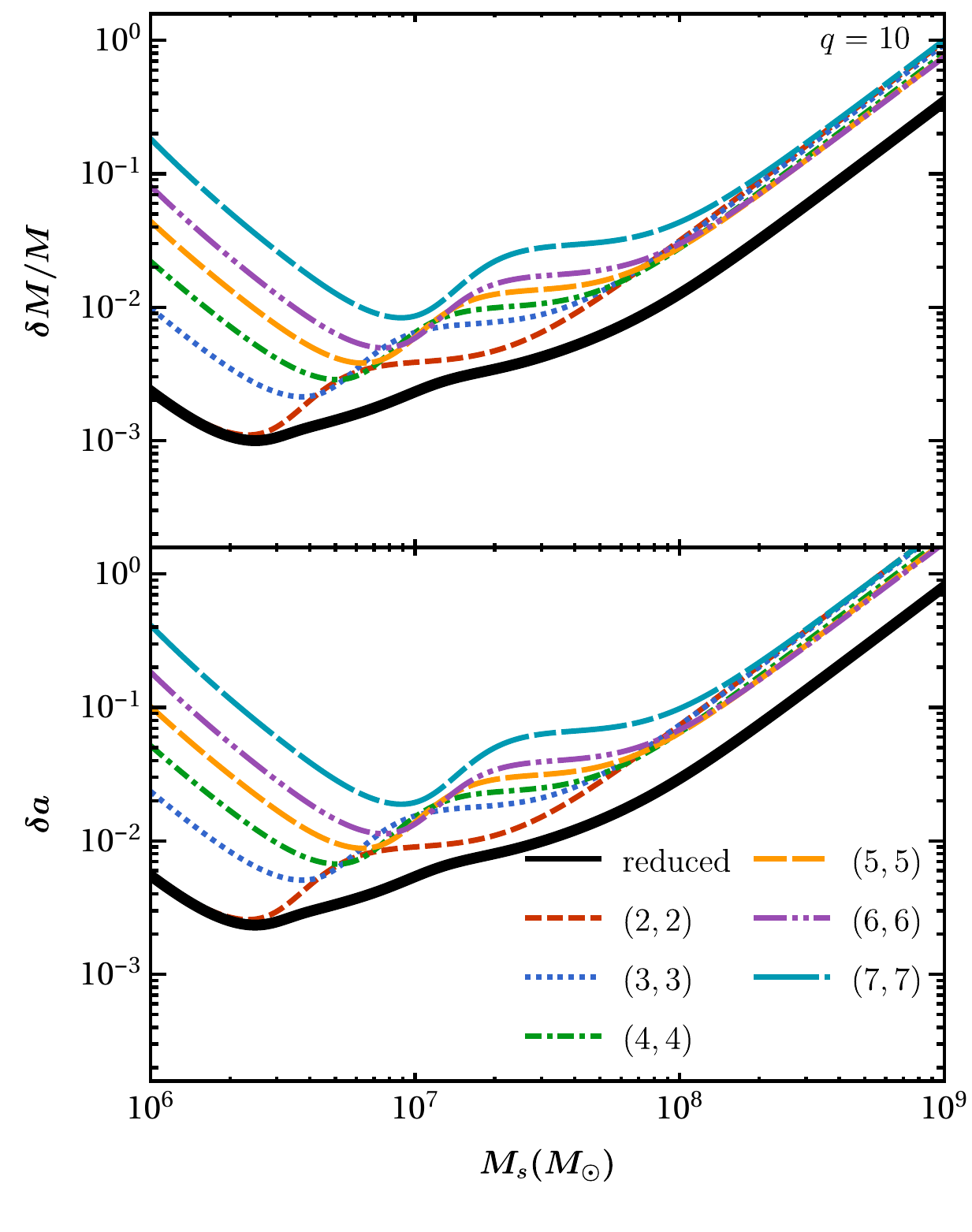}
  \end{tabular}
  \caption{Angle-averaged errors on the remnant's redshifted mass (top panel) and dimensionless spin (bottom panel) as a function of the remnant's total mass. We consider a binary merger of mass ratio $q=2$ (left) and $q=10$ (right) at $z=1$. Each line corresponds to a different mode; the thick, solid black line corresponds  to the total error obtained after combining all modes.} 
\label{fig:errRemnantz1}
\end{figure*}

\begin{figure*}[t]
\begin{tabular}{cc}
  \includegraphics[width=0.48\textwidth]{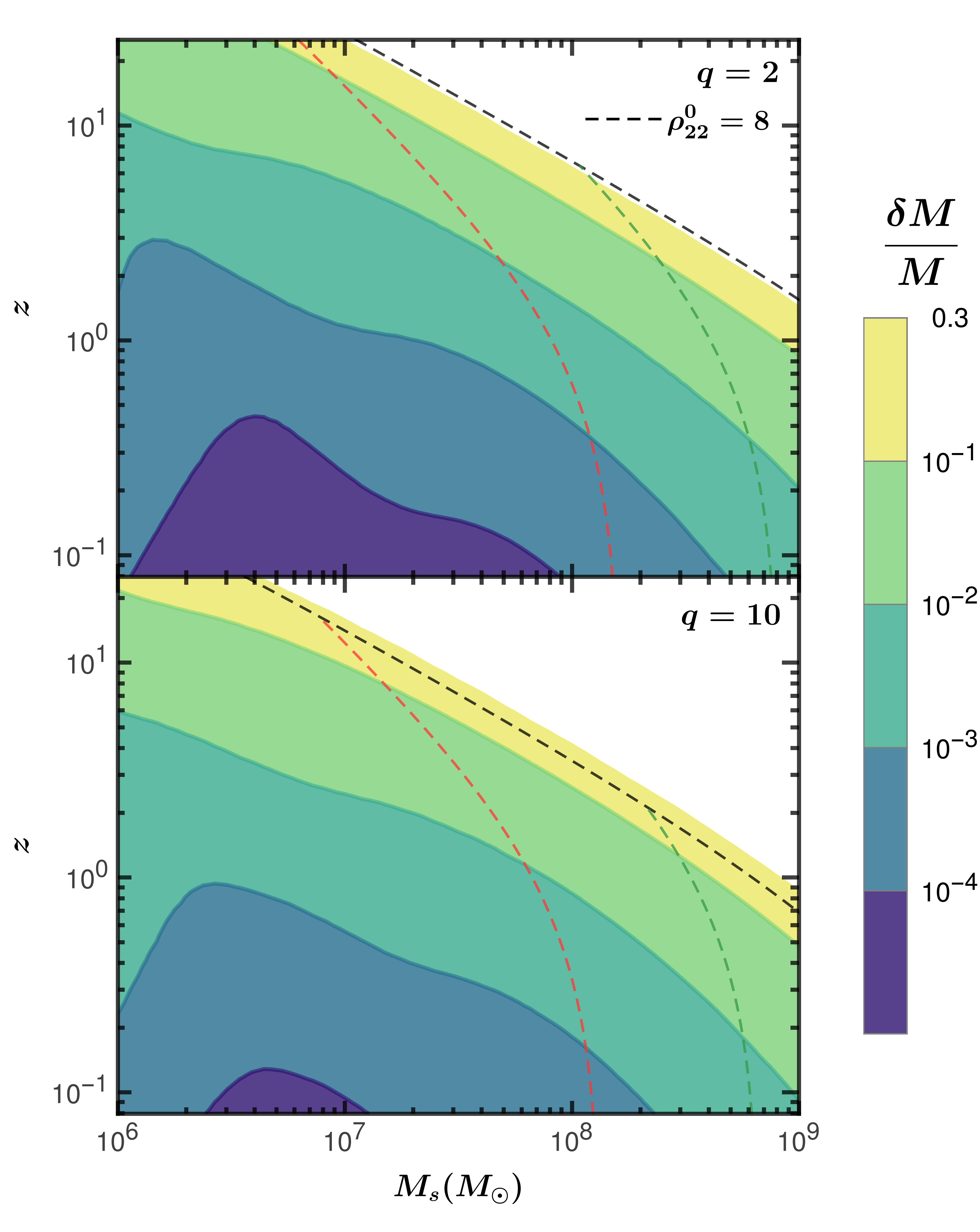}&
  \includegraphics[width=0.48\textwidth]{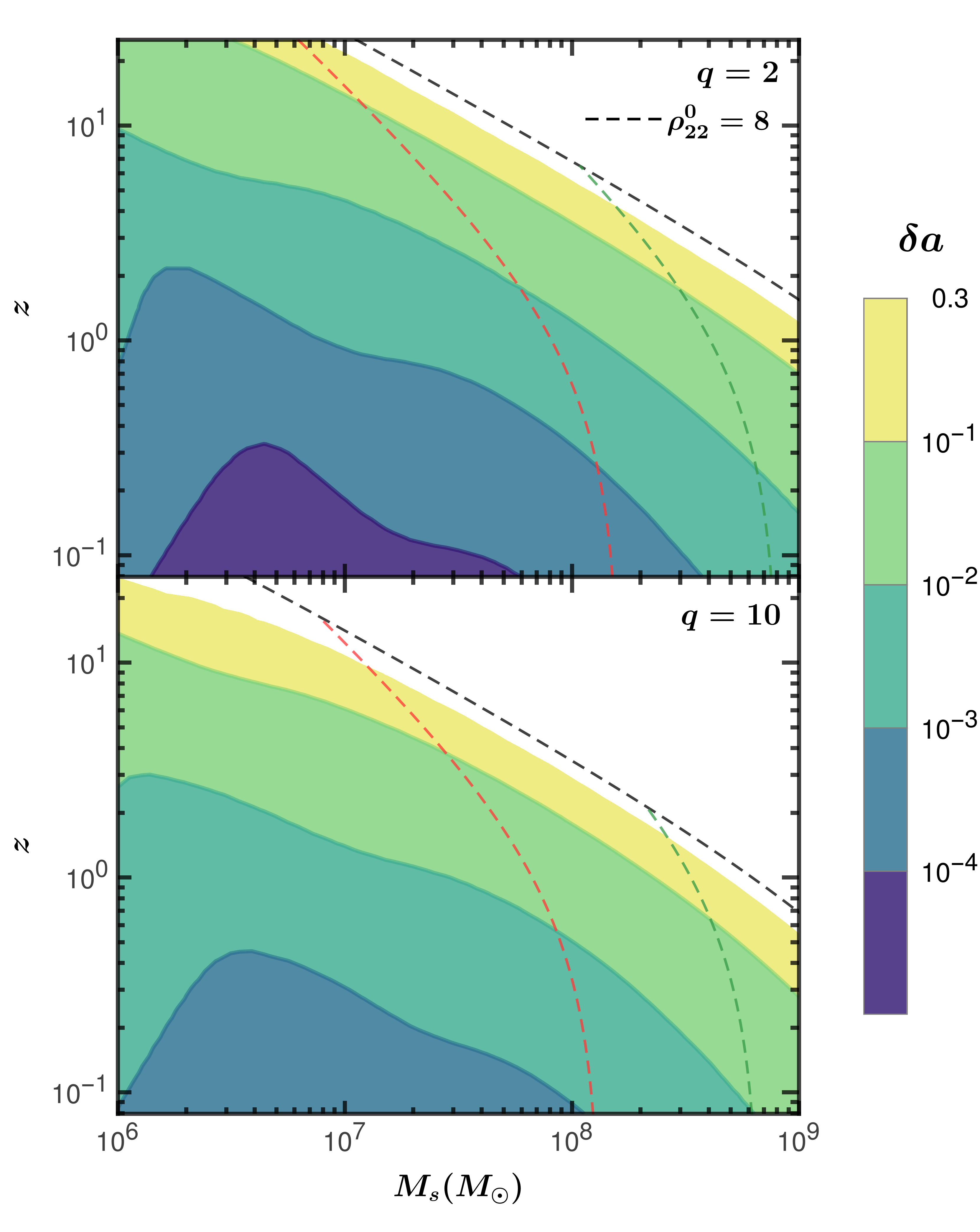}
  \end{tabular}
  \caption{Median relative error on the detector-frame mass (left) and median absolute error on the remnant spin (right) for binary mergers with $q=2$ (top panels) and $q=10$ (bottom panels). We also show the horizon of $(2,\,2)$ mode and redshifted-mass cutoff at $f_{\rm cut}=10^{-4}$~Hz (in red) $f_{\rm cut}=2\times10^{-5}$~Hz (in green). %
  } 
\label{fig:errRemContours}
\end{figure*}

\section{Remnant mass and spin}
\label{sec:remnant}

For our present purposes we can model the LISA detector in the low-frequency approximation as a combination of two independent LIGO-like detectors or ``channels'' (denoted by a superscript $i=$``I'' or ``II'') with antenna pattern functions $F_{+\,,\times}^{\rm I, II}$ and sky-sensitivities $\Omega_{\ell m}^{\rm I, II}$~\cite{Cutler:1997ta,Berti:2004bd}. %
The ringdown signal from a BH with source-frame mass $M_{s}$, redshifted mass $M=M_{s}(1+z)$ and dimensionless spin $a$ measured by each detector can be written in the time domain as a superposition of damped sinusoids of the form
\be\label{eq:hdef}
h_{\ell m}^i(t) =  {\mathcal A}^i_{\ell m} e^{-(t-t_0)/\tau_{\ell m}}
\cos\left (2\pi f_{\ell m} t + \Phi^i_{\ell m}\right ),
\ee
where $f_{\ell m}=f^{(s)}_{\ell m}/(1+z)$ is the redshifted (detector-frame) frequency, $\tau_{\ell m}=\tau^{(s)}_{\ell m}(1+z)$ is the redshifted decay time, and for later convenience we also define the quality factor $Q_{\ell m}=\pi f_{\ell m}\tau_{\ell m}$.

The signal phase $\Phi^i_{\ell m}$ is given by
\be\label{eq:phidef}
\Phi^i_{\ell m}  =  \phi_{\ell m}-2\pi f_{\ell m} t_0 + m\, \varphi + \tan^{-1} \left(\frac{F^i_\times\, Y^{\ell m}_\times}
{F^i_+\, Y^{\ell m}_+} \right),\,
\ee
where $t_0$ is the starting time of the signal.

The signal amplitude in the $i$-th detector is
\be\label{eq:Adef}
{\mathcal A}^i_{\ell m}=\frac{M \Omega^i_{\ell m}}{d_L} A_{\ell m}(q)\,,
\ee
where $d_L=d_L(z)$ is the luminosity distance to the source (we use the standard cosmological parameters determined by \emph{Planck}~\cite{Ade:2015xua}),
\begin{equation}
  \Omega^i_{\ell m}\equiv \sqrt{\left(F^i_+ Y ^{\ell m}_+\right)^2+\left(F^i_\times Y ^{\ell m}_\times\right)^2}
  \label{eq:Omegalm}
\end{equation}
is a ``sky sensitivity'' coefficient and $A_{\ell m}$ is a ringdown
excitation amplitude, which depends on the mass ratio of the binary
and on the spins of the
progenitors~\cite{London:2014cma,London:2018gaq,Kamaretsos:2011um,Baibhav:2018rfk,Berti:2007fi}. We
compute $A_{\ell m}$ as described in Ref.~\cite{Baibhav:2018rfk}. We
consider only nonspinning binaries and we neglect precession
(cf.~\cite{Lim:2019xrb,Apte:2019txp,Hughes:2019zmt} for a calculation
of ringdown excitation amplitudes of more general trajectories in the
extreme mass-ratio limit).

The antenna pattern functions $F^i_{+,\times}$ depend on the
source sky position angles $(\theta,\phi)$ and on the polarization
angle $\psi$~\cite{Cutler:1997ta}:
\bea
F^{\rm I}_+ (u,\phi,\psi)&=&\frac{1+u^2}{2} \cos{2 \psi} \cos{2 \phi}-u \sin{2 \psi } \sin{2 \phi }\,,\nonumber\\
F^{\rm I}_\times (u,\phi,\psi)&=&\frac{1+u^2}{2}  \sin{2 \psi} \cos{2 \phi}+u \cos{2 \psi} \sin{2 \phi }\,,\nonumber\\
F^{\rm II}_{+,\times}(u,\phi,\psi) &=& F^{\rm II}_{+,\times}\left(u,\phi-\f{\pi}{4},\psi\right)\,,
\eea
where $u=\cos{\theta}$.
The harmonics $Y^{\ell m}_{+,\times}$ corresponding to the two ringdown polarizations can be found by summing over modes with positive and negative $+m$, as follows~\cite{Berti:2007zu,Kamaretsos:2011um,Kamaretsos:2012bs}:
\bea
Y^{\ell m}_+(\iota) & \equiv & {_{-2}Y^{\ell m}}(\iota,0) + (-1)^\ell\,  {_{-2}Y^{\ell -m}}(\iota,0),\nonumber\\
Y^{\ell m}_\times(\iota) & \equiv & {_{-2}Y^{\ell m}}(\iota,0) - (-1)^\ell\,  {_{-2}Y^{\ell -m}}(\iota,0).
\eea
Here $\iota$ is the angle between the spin axis of the remnant and the plane of the sky. For example, for $\ell=m=2$ we get %
\bea
Y^{22}_+(\iota)&=&\frac{1}{4} \sqrt{\frac{5}{\pi }} \left[1+(\cos\iota)^2\right]\,, \nn\\ 
Y^{22}_\times(\iota)&=&\frac{1}{2} \sqrt{\frac{5}{\pi }} \cos\iota\,.
\label{Ypc22}
\eea

Ref.~\cite{Berti:2005ys} used a Fisher matrix analysis to estimate errors on the detector amplitude ${\mathcal A}^i_{\ell m}$ and on the phase $\Phi ^i_{\ell m}$:
\bea
\frac{\delta {\mathcal A}^i_{\ell m}}{{\mathcal A}^i_{\ell m}}=\frac{\sqrt{2}}{\rho_{\ell m}^i}\,,\\
\delta  \Phi_{\ell m}^i=\frac{1}{\rho_{\ell m}^i}\,.
\eea
Here $\rho_{\ell m}^i$ denotes the signal-to-noise ratio (SNR) in detector $i$~\cite{Baibhav:2018rfk}:
\be
\label{eq:rho0}
\rho_{\ell m}^i= \rho_{\ell m}^0 w_{\ell m}^i (\iota,\theta,\phi,\psi)\,,
\ee
where $\rho_{\ell m}^0$ is a detector-independent optimal SNR, while $ w_{\ell m}^i (\iota,\theta,\phi,\psi)= \Omega^i_\lm/\max(\Omega^i_\lm)\le 1$ is a ``projection factor'' that depends on the sky location, inclination and polarization angles.

Ref.~\cite{Berti:2005ys} also showed that a quasinormal mode with signal-to-noise ratio (SNR) $\rho_\lm=\left[(\rho_\lm^{\rm I})^2+(\rho_\lm^{\rm II})^2\right]^{1/2}$ can be used to measure the redshifted mass and spin of the remnant with accuracy
\bea
\delta a &=& \frac{1}{\rho_{\ell m}}\left|2\frac{Q_{\ell m}}{Q_{\ell m}'}\right|\,,\\
 \frac{\delta M}{M} &=& \frac{1}{\rho_{\ell m}}\left|2\frac{Q_{\ell m}
f_{\ell m}'}{Q_{\ell m}'f_{\ell m}}\right|\,,
\label{eq:mass-spin}
\eea
which is independent of the channel, since we are summing over $i=\rm I, II$.
In other words, the error $\sigma$ resulting from two-detector measurements is $\sigma^{-2}=\sigma_{\rm I}^{-2}+\sigma_{\rm II}^{-2}$, which is equivalent to replacing the SNR $\rho_\lm^{i}$ in each detector by the total SNR $\rho_\lm$.
Therefore, in this section and in the next we will drop the subscript $i$.

Estimates of mode excitation based on numerical relativity simulations suggest that, in favorable cases, LISA may see all multipolar components of the radiation that have been computed in current numerical relativity simulations~\cite{Baibhav:2018rfk}. Parameter estimation errors could be further reduced for these ``golden binaries'', as we show in Fig.~\ref{fig:errRemnantz1}. We consider a binary with $q=2$ (left panels) and $q=10$ (right panels) at $z=1$ and we plot angle-averaged parameter estimation errors on redshifted mass and spin inferred from specific modes, as well as the (smaller) total error estimate when we consider all multipoles. We assume Gaussian distributions for the errors from each mode, and we estimate the total error as~
\bea\label{eq:redError}
\left(\f{\delta M}{M}\right)_{\rm reduced}^{-2} &=& \sum _{\ell m} \left(\f{\delta M}{M}\right)_{\lm}^{-2}\,,\\
\left(\delta a_{\rm reduced}\right)^{-2} &=& \sum _{\ell m} \left(\delta a_{\lm}\right)^{-2}\nn\,,
\eea
where $\left(\delta M/M\right)_{\lm}$ is the relative error on the remnant's redshifted mass and $\delta a_{\lm}$ is the absolute error on its dimensionless spin computed using the $(\ell,\,m)$ mode. For small mass ratios most of the parameter estimation accuracy comes from the $(2,\,2)$ mode, while higher multipoles make almost no contribution to the total error. The scenario changes drastically for $q=10$: now all harmonics have SNR comparable to that of the $(2,\,2)$ mode, the errors from the individual modes are comparable, and adding them in quadrature leads to a significant improvement in parameter estimation.

In Fig.~\ref{fig:errRemContours} we show contour plots for the median relative error $\delta M/M$ on the redshifted mass (left) and for the median absolute error $\delta a$ on the dimensionless spin (right).

LISA can measure BH remnant spins for binaries with $q=2\,(10)$ with an accuracy of $0.01$ up to redshift $z=9.8\,(2.6)$ if $M=10^6 M_\odot$, or up to redshift $z\approx 1.2\,(0.5)$ if  $M\sim 10^8 M_\odot$. LISA can also measure the redshifted mass of the remnant for binaries with $q=2\,(10)$ with an accuracy of $1\%$ up to redshift $z=12\,(6)$ if $M=10^6 M_\odot$, or up to redshift $z\approx 1.5\,(0.8)$ if  $M\sim 10^8 M_\odot$.

Interestingly, the remnant spins and  redshifted masses for binaries with $q=2\,(10)$ can be measured with an accuracy of $10\%$ even if the remnant has mass as large as $M\sim 10^9 M_\odot$, as long as the merger occurs at $z<0.7\,(0.3)$. Such binaries are usually thought to be observable only with Pulsar Timing Arrays (PTAs). It is possible that PTAs may observe the early inspiral of a few resolvable binaries with $z<1$~\cite{Sesana:2008xk}, while LISA may observe their merger-ringdown.

\section{Mass ratio and inclination}
\label{sec:qiota}

In this section we will exploit the fact that the excitation of different modes with $\ell=m$ depends in a characteristic way on the mass ratio $q$ and on the inclination angle $\iota$ to infer $q$ and $\iota$.
Let us focus first on one of the two independent LIGO-like detectors, dropping the superscripts (I, II) for clarity.

\begin{figure}[t]
  \includegraphics[width=\columnwidth]{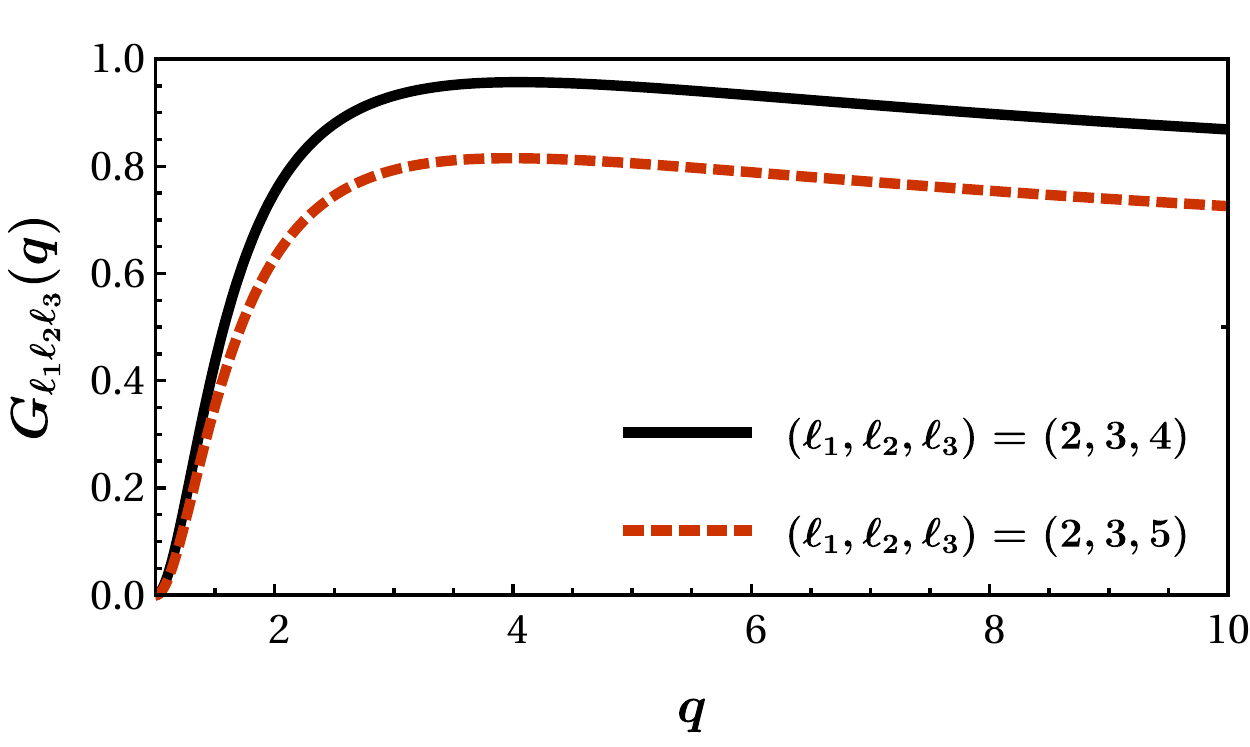}
  \caption{The function $G_{\ell_1\ell_2\ell_3}(q)$ defined in Eq.~(\ref{eq:Gdef}).} 
\label{fig:G}
\end{figure}

For multipoles with $\ell=m$, the sky sensitivity appearing in Eq.~(\ref{eq:Adef}) is of the form
\be
\Omega_{\ell \ell}=r_{\ell} (\sin\iota)^{\ell-2} \Omega_{22}\,,
\label{eq:Omegall}
\ee
where the proportionality constant
\be
r_\ell= \frac{(-1)^\ell 2^{2-\ell}}{\sqrt{5}} \sqrt{\frac{(2 \ell)!(2 \ell+1)}{(\ell-2)! (\ell+2)!}}\,
\ee
is such that
\be
Y_{+,\times}^{\ell\ell} =r_\ell (\sin\iota)^{\ell-2} Y_{+,\times}^{22}\,.
\label{Ypcrossll}
\ee
The detector-amplitude ratio of two modes -- which to simplify the notation we shall denote as, say, ${\mathcal A}_{\ell_i}={\mathcal A}_{\ell_i m_i}$  with $\ell_i=m_i$ -- depends only on $q$ and $\iota$, i.e.
\be \label{eq:ampRatio1}
\frac{{\mathcal A}_{\ell_2}}{{\mathcal A}_{\ell_1}}=(\sin \iota)^{\ell_2-\ell_1} H_{\ell_1 \ell_2}(q)\,,
\ee
where 
\be
H_{\ell_1 \ell_2}(q)\equiv \frac{r_{\ell_2}}{r_{\ell_1}} \frac{A_{\ell_2}(q)}{A_{\ell_1}(q)}\,.
\ee
By a simple extension, we can obtain a three-mode combination which depends only on $q$:
\be\label{eq:ampRatio2}
\frac{\mathcal{A}_{\ell_2}^\lambda}{\mathcal{A}_{\ell _3} \mathcal{A}_{\ell_ 1}^{\lambda-1}}= G_{\ell_1 \ell_2 \ell_3}(q)\,,
\ee
where
\be\label{eq:Gdef}
 G_{\ell_1 \ell_2 \ell_3}(q) \equiv \frac{r_{\ell_2}^\lambda}{r_{\ell_3} r_{\ell_1}^{\lambda-1}} \frac{A_{\ell_2}^\lambda}{A_{\ell _3} A_{\ell_ 1}^{\lambda-1}}
\ee
and $\lambda\equiv (\ell_3-\ell_1)/(\ell_2-\ell_1)$.  This function is
plotted in Fig.~\ref{fig:G} in two cases of interest:
$(\ell_1,\,\ell_2,\,\ell_3)=(2,\,3,\,4)$ and
$(\ell_1,\,\ell_2,\,\ell_3)=(2,\,3,\,5)$. Note that
$G_{\ell_1 \ell_2 \ell_3}(q)$ has a local maximum for $q\sim 4$ in both
cases. This observation will be useful later.

Note that $G_{\ell_1 \ell_2 \ell_3}(q)$ is obtained by fitting
ringdown excitation amplitudes to numerical simulations. Higher
harmonics are typically subdominant and contaminated by numerical
noise. Since the errors are proportional to
$G_{\ell_1 \ell_2 \ell_3}(q)/G_{\ell_1 \ell_2 \ell_3}'(q)$, our
results are very sensitive to the accuracy of these fits (and
therefore, indirectly, to the accuracy of the numerical
simulations). This is why we do not use modes with $\ell>5$ to
estimate $q$ and $\iota$, even though those modes were used to
estimate $M$ and $a$.

\begin{figure*}[t]
\begin{tabular}{cc}
  \includegraphics[width=0.492593\textwidth]{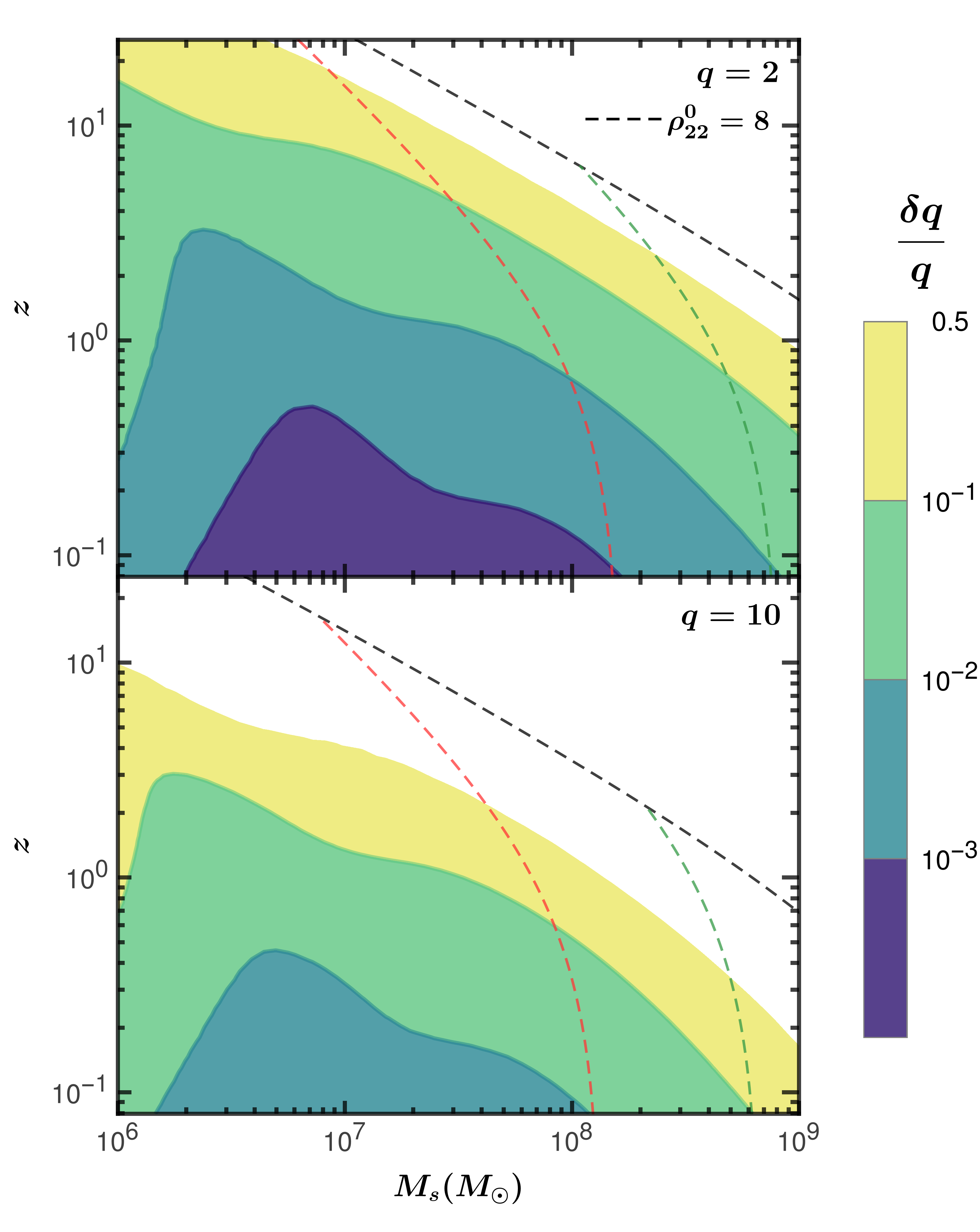}
  \includegraphics[width=0.507407\textwidth]{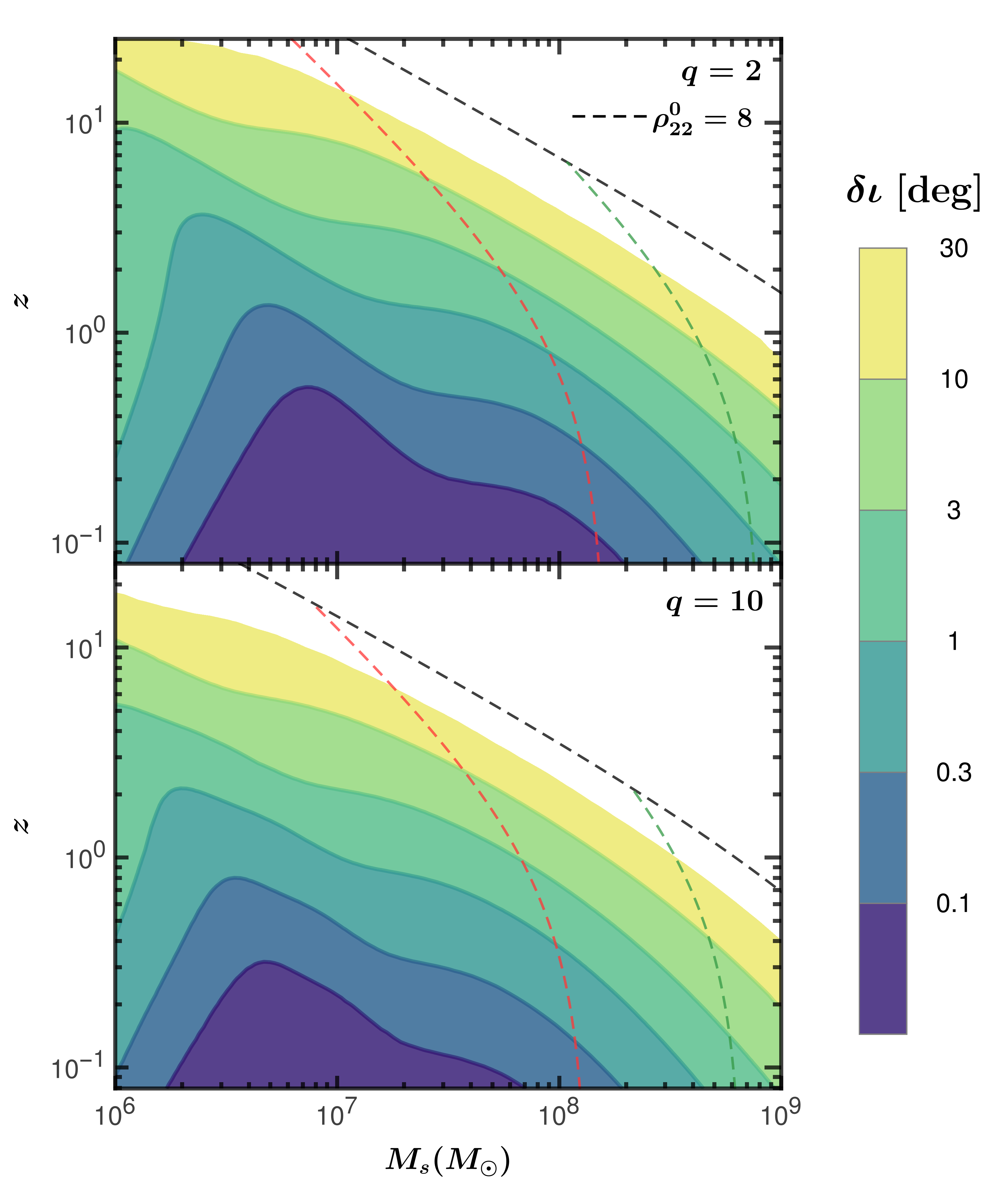}
\end{tabular}
  \caption{Median relative errors $\delta q/q$ for the mass ratio (left) and median error on the inclination angle $\iota$ (right) for nonspinning binary mergers with mass ratio $q=2$ (top) and $q=10$ (bottom). 
We also show the horizon of $(2,\,2)$ mode and redshifted-mass cutoff at $f_{\rm cut}=10^{-4}$~Hz (in red) $f_{\rm cut}=2\times10^{-5}$~Hz (in green). %
  }
\label{fig:errContoursQI}
\end{figure*}

The idea is now to infer $q$ and $\iota$ from the detector amplitudes ${\mathcal A}_{\ell_i}$ of the three dominant modes. To estimate measurement errors on $q$ and $\iota$, we propagate errors from the basis $\{{\mathcal A}_{\ell_1},\,{\mathcal A}_{\ell_2},\,{\mathcal A}_{\ell_3}\}$ to the basis $\{q,\,\iota\}$ as follows:
\be
\cov_{\ell_1\ell_2\ell_3}(q,\iota)=\jacob{(q,\,\iota)}{{\mathcal A}_{\ell_1\ell_2\ell_3}}\cdot\cov({\mathcal A})\cdot\left(\jacob{ (q,\,\iota)}{{\mathcal A}_{\ell_1\ell_2\ell_3}}\right)^T\,,
\ee
where  $\cov({\mathcal A})$ is the diagonal covariance matrix of detector amplitudes with elements $2\f{{\mathcal A}_{\ell}^2}{\rho_{\ell}^2}$, and $\jacob{(q,\,\iota)}{{\mathcal A}_{\ell_1\ell_2\ell_3}}$ denotes the Jacobian of the transformation between the two bases, obtained from Eqs.~(\ref{eq:ampRatio1}) and (\ref{eq:ampRatio2}). We can also use multiple mode combinations to reduce the uncertainty:
\be\label{eq:covqired}
\cov(q,\iota)^{-1}=\sum_{\{ \ell_1,\ell_2,\ell_3\}}\cov_{\ell_1\ell_2\ell_3}(q,\iota)^{-1}\,.
\ee

The left panel of Fig.~\ref{fig:errContoursQI} shows contour plots of the median relative error on the mass ratio $\delta q/q$ (left) for sources uniformly distributed over the sky.  To reduce the error we follow the procedure outlined above, using the following two combinations of modes: $(\ell_1,\,\ell_2,\,\ell_3)=(2,\,3,\,4)$ and $(3,\,4,\,5)$. The top panels show that for a binary with mass ratio $q=2$, LISA can measure $q$ with an accuracy of $10\%$ up to redshift $z=16 (2.1)$ for BHs of mass $10^6 M_\odot (10^8 M_\odot)$. In the bottom panels we consider a binary with mass ratio $q=10$, and we show that measuring the mass ratio is harder: in this case we can get $q$ with better than $10\%$ accuracy out to $z=0.7$ for $M_s=10^6 M_\odot$.
The right panel of Fig.~\ref{fig:errContoursQI} shows median error contours for the inclination angle $\iota$. For a  $q=2$ binary (top panel), LISA can measure $\iota$ within $10\degree$ up to $z\approx18\,(2.4)$ for BHs of mass $10^6 M_\odot\,(\sim 10^8 M_\odot)$. In the bottom panel we consider a $q=10$ binary, for which $q$ is harder to measure, but the inclination can still be measured to a relatively good accuracy: we can measure $\iota$ within $10\degree$ up to redshift $z\approx 11\,(1.4)$ for BHs of mass $10^6 M_\odot\,(\sim 10^8 M_\odot)$. The dependence of the various errors on the binary parameters will be discussed in more detail in Sec.~\ref{sec:qIotaSkyDep} below.

\section{Sky localization}
\label{sec:skyloc}

In general, LISA can localize inspiraling sources and measure their distance by using amplitude and phase modulations due to the orbital motion of the constellation around the Sun~\cite{Cutler:1997ta,Vecchio:2003tn,Berti:2004bd,Lang:1900bz,Lang:2007ge}.  This is not possible when we observe only the merger/ringdown, because then the signal duration is very short: even for remnant masses as large as $\sim 10^9 M_{\odot}$ the signal can last at most $\sim 17$~hours, compared to the LISA orbital time scale $T\sim 1$~yr.\footnote{In principle, for such massive binaries we could still measure first-order corrections to the antenna pattern due to orbital modulations.  However, in Appendix~\ref{sec:evolutionLocalization} we show that these modulations can be measured with a typical accuracy $\propto T/\tau_{22}$, which is not sufficient even for the most massive remnants.}
For this reason we will explore other ways of localizing the source, which are mainly based on comparing the amplitudes and phases of the harmonics measured in different channels.

\subsection{Localization contours using the amplitudes and phases of the dominant mode in different channels}

\begin{figure}[t]
  \includegraphics[width=\columnwidth]{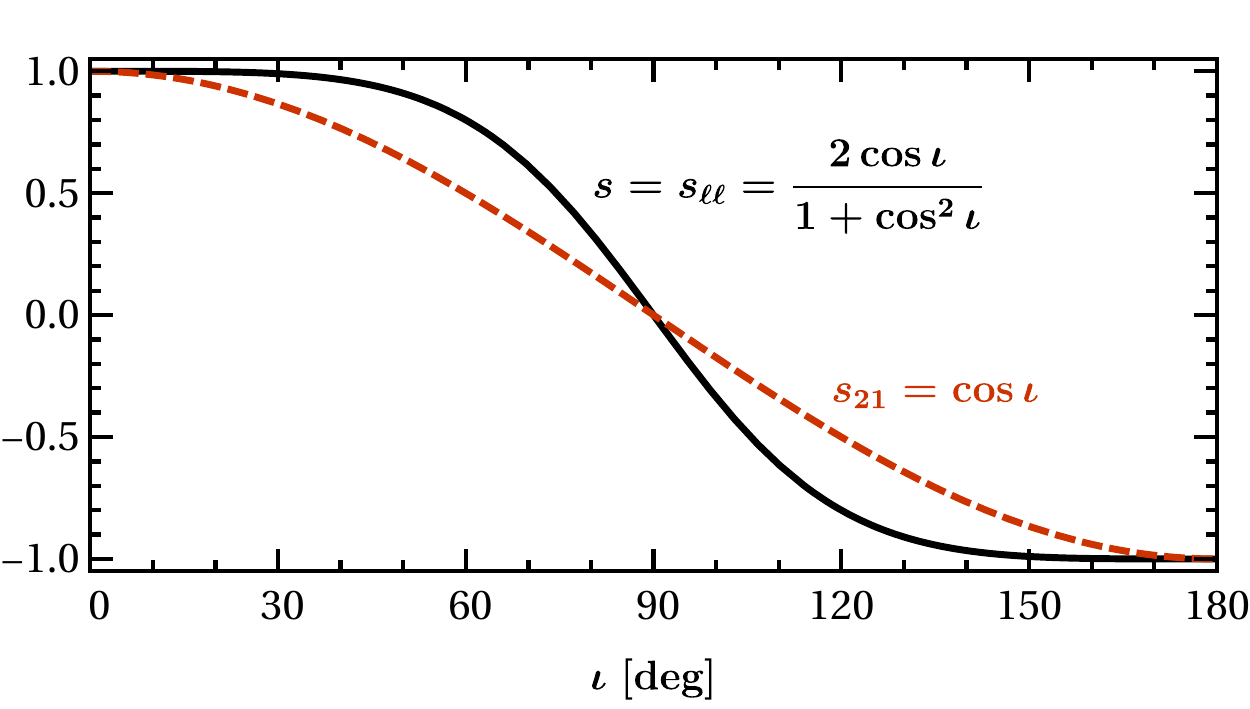}
  \caption{The function $s=s_{\ell \ell}$ [cf. Eq.~(\ref{sll})] and the function $s_{21}$.} 
\label{fig:slm}
\end{figure}

A first possibility to determine the sky location of a source is to take the ratio of the signal amplitudes in two channels %
\begin{align}\label{eq:QAdef}
Q_A^{\ell m}=\left(\f{{\mathcal A}^{\rm I}_{\ell m}}{{\mathcal A}^{\rm II}_{\ell m}}\right)^2
&=\left(\f{\Omega^{\rm I}_{\ell m}(\iota,\theta,\phi,\psi)}{\Omega^{\rm II}_{\ell m}(\iota,\theta,\phi,\psi)}\right)^2 \nn\\
&=\f{\left(F^{\rm I}_+\right)^2+s_{\ell m}^2\left(F^{\rm I}_\times\right)^2}{\left(F^{\rm II}_+\right)^2+s_{\ell m}^2\left(F^{\rm II}_\times\right)^2}
\end{align}
and the difference of the phases 
 measured in the two channels
\begin{align}\label{eq:QPdef}
\tan^{-1}Q_\Phi^{\ell m}&=\Phi^{\rm II}_{\ell m}-\Phi^{\rm I}_{\ell m} \nn\\  
&= 
 \tan^{-1} \left(\frac{s_{\ell m} F^{\rm II}_\times}
{F^{\rm II}_+} \right)- \tan^{-1} \left(\frac{s_{\ell m} F^{\rm I}_\times}
{F^{\rm I}_+} \right).
\end{align}
where we have defined the function $s_{\ell m}(\iota)=Y^{\ell m}_\times(\iota)/Y^{\ell m}_+(\iota)$, and we have omitted the inclination dependence for brevity.
From Eqs.~(\ref{Ypc22}) and (\ref{Ypcrossll}) it follows that
\begin{equation}
  s=s_{\ell\ell}=s_{22}=\frac{2 \cos\iota}{1+\cos^2\iota}
  \label{sll}
\end{equation}
for all modes with $\ell=m$. This function is plotted in Fig.~\ref{fig:slm} along with the corresponding function $s_{21}$.

The amplitude ratio $Q_A=Q_A^{22}$ and phase difference $Q_\Phi=Q_\Phi^{22}$ of the dominant mode with $\ell=m=2$ are the two main observable quantities. Let us assume that we have determined the inclination $\iota$ as described in Sec.~\ref{sec:qiota}. Then the two observables $(Q_A, Q_\Phi)$ depend on three unknowns ($\theta$, $\phi$ and $\psi$). Since, at this stage, this system is underdetermined 
 we cannot find the exact sky location $(\theta, \phi)$, but we can infer {\em contours} of constant $(Q_A,\,Q_\Phi)$ in the sky.

 For the moment we will ignore measurement errors on $Q_A$ and $Q_\Phi$, which scale like $1/\rho_{22}$. This assumption is justified: the limiting factor in the measurement is the inclination $\iota$, determined (as we discussed previously) from subdominant modes such as $(\ell,\,m)=(4,\,4)$ or $(5,\,5)$, which typically have smaller signal-to-noise ratio than the $(2,\,2)$ mode.

\begin{figure}[htp]
\begin{tabular}{c}
\includegraphics[width=0.9\columnwidth]{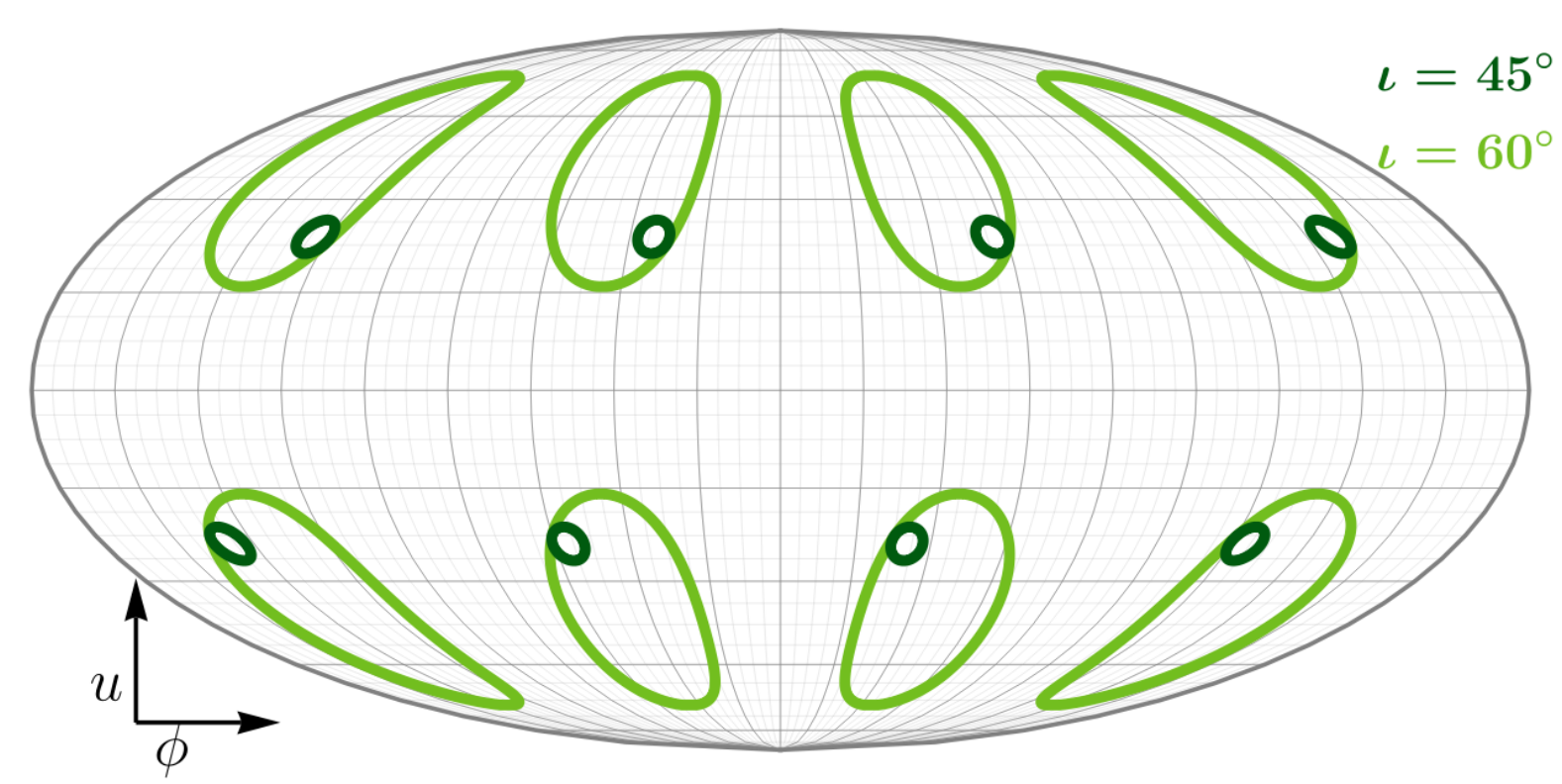}\\
\includegraphics[width=0.9\columnwidth]{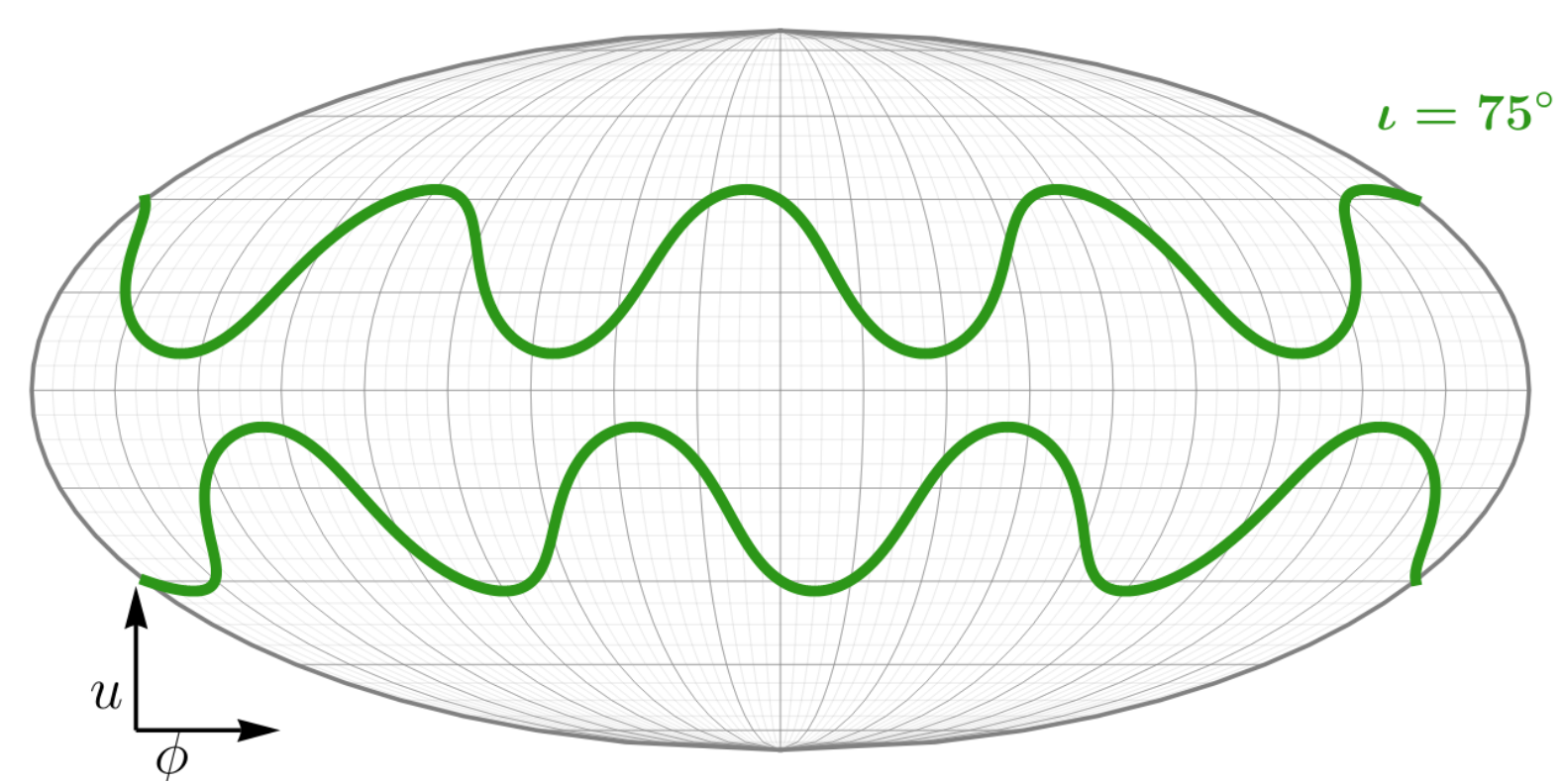} \\
\includegraphics[width=0.9\columnwidth]{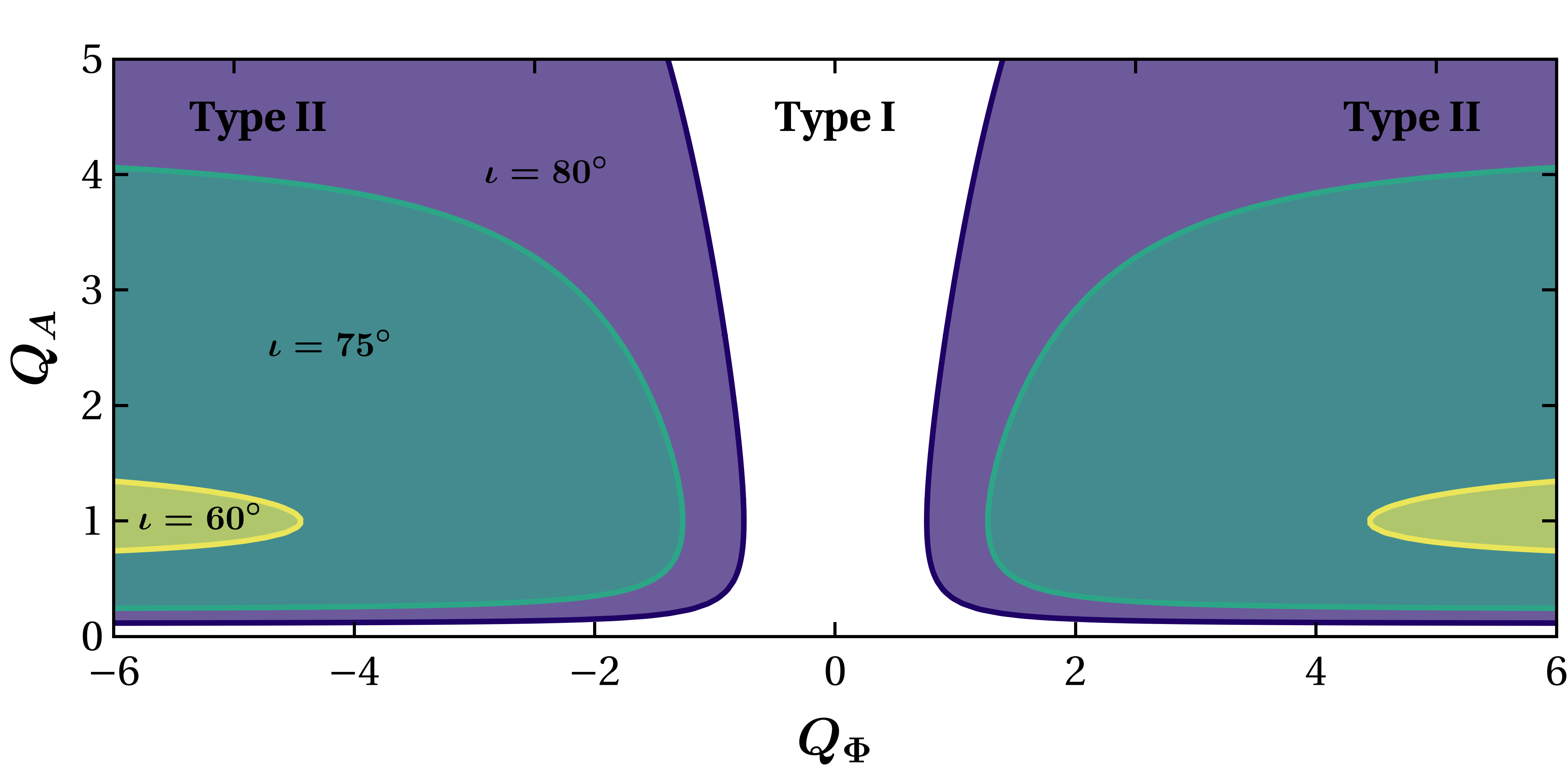}
\end{tabular}
\caption{Top and central panels: localization contours found using relative detector amplitudes $Q_A$ and phases $Q_\Phi$ for the dominant $(2,\,2)$ mode. Here we consider a source at $(u=\cos\iota,\,\phi,\,\psi) = (0.5,\,60\degree,\,60\degree)$  and three selected values of the inclination: $45\degree$ and $60\degree$ (top panel) and $75\degree$ (central panel). For smaller inclinations ($\iota=45\degree$ and $\iota=60\degree$) we get Type I contours, according to the definition in the main text. For larger inclinations ($\iota=75\degree$) we get Type II contours. The bottom panel shows a phase diagram of the different classes of solutions in the $(Q_A, Q_\Phi)$ plane for three fixed values of the inclination.%
} 
\label{fig:QAP}
\end{figure}

 By eliminating $\psi$ from Eqs.~(\ref{eq:QAdef}) and~(\ref{eq:QPdef}) we get contours in the $(\theta,\,\phi)$ plane. These belong to two classes of solutions, as illustrated in Fig.~\ref{fig:QAP}: 

\begin{itemize}
\item{{\em Type I:}} the contours form a set of $8$ closed rings, and there can be anywhere from $0$ to $4$ solutions at a given $\phi$ (top panel of Fig.~\ref{fig:QAP}).

\item{{\em Type II:}} the contours form two ring-like structures enclosing the north and south pole, and there are two solutions at any given $\phi$  (middle panel of Fig.~\ref{fig:QAP}).
\end{itemize}

These two classes of solution arise because the equations have a different number of solutions in different regions of the $(Q_A, Q_\Phi)$ parameter space: ring-like solutions of Type II arise when
\be\label{eq:typeSep}
s^2<Q_A<\frac{1}{s^2}\;{\rm and}\; \left| Q_\Phi\right| >\frac{(Q_A+1) \left| s\right|}{\sqrt{\left(Q_A-s^2\right) \left(1-Q_A s^2\right)}}.
\ee

In the bottom panel of Fig.~\ref{fig:QAP} we plot the ``phase diagram'' of solutions in the $(Q_A, Q_\Phi)$ parameter space for a source at $\theta=\phi=\psi=60\degree$ and three fixed values of $\iota=45\degree,\,60\degree,\,75\degree$. 
Type II solutions are usually present for nearly edge-on binaries. Most of the solutions are of Type I, with only about $1/4$ of sources belonging to Type II if we assume that they are isotropically distributed.
Notice also that the rings are symmetric under parity ($u\equiv \cos \theta\to-u,\, \phi\to 2\pi -\phi$).

In practice, the rings will have finite ``widths'' which are mainly determined by the uncertainty in $\iota$.

The discussion above focused on modes with $\ell=m$, but it is also applicable to $\ell\neq m$ modes, with the $(2,\,1)$ mode being the most relevant observationally. The main difference is that $s_{21}=\cos\iota$. The $(2,\,1)$ mode also yields two families of solutions, with the ``phase diagram'' being determined by Eq.~\ref{eq:typeSep}. The three Type II regions shown in Fig.~\ref{fig:QAP} -- which correspond to $\iota=60\degree,\,75\degree,\,80\degree$ for the $(2,\,2)$ mode -- would correspond to $\iota=36.9\degree,\,61\degree,\,70.3\degree$ for the $(2,\,1)$ mode. In other words, Type II solutions are more likely for the $(2,\,1)$ mode: about one third of the sky gives Type II solutions for the $(2,\,1)$ mode, compared to about one fourth of the sky for the $(2,\,2)$ mode.

\subsection{Localization contours using the amplitude of the $(2, 1)$ mode}
\label{sec:s21}

In the previous section we inferred localization contours using the amplitude ratio $Q_A=Q_A^{22}$ and phase difference $Q_\Phi=Q_\Phi^{22}$ of the dominant mode with $\ell=m=2$, assuming that the inclination has been measured as described in Sec.~\ref{sec:qiota}. Unfortunately we cannot extract any more information from the remaining modes with $\ell=m$, because the sky sensitivity $\Omega_{\ell \ell}\propto \sin(\iota)^{\ell-2} \Omega_{22}$ for all of these modes: cf. Eq.~(\ref{eq:Omegall}).

More information on the pattern functions $F^i_{+,\,\times}$ is encoded in modes with $\ell\neq m$. The excitation of these modes is generally harder to quantify through numerical relativity simulations, where subdominant modes are usually contaminated by dominant modes through a mixing of spherical and spheroidal harmonics with the same $m$ and lower $\ell$~\cite{Berti:2005gp,Buonanno:2006ui,Berti:2007fi,Kelly:2012nd,London:2014cma,Berti:2014fga}. The $(2,\,1)$ mode is an exception, because (i) it is not affected by mode mixing, and (ii) it can be excited to relatively large amplitudes, especially for spinning BH binaries~\cite{Berti:2007nw,Kamaretsos:2011um,Kamaretsos:2012bs,London:2014cma,Baibhav:2017jhs,Baibhav:2018rfk}.

In this section we will focus on the localization information contained in the $(2,\,1)$ mode. Let us assume that the inclination angle $\iota$ and the mass ratio $q$ are known. Then a possible strategy would be to think about the two sky sensitivities $(\Omega^i_{22}, \Omega^i_{21})$ (or more precisely, the corresponding measurable detector amplitudes ${\mathcal A}^i_{\ell m}\propto \frac{\Omega^i_{\ell m}}{d_L}$) as functions of the
corresponding antenna pattern functions $(F^i_{+}, F^i_{\times})$ in each channel [cf. Eq.~(\ref{eq:Omegalm})], and to solve these equations to determine $(F^i_{+}, F^i_{\times})$ in each channel. A problem with this strategy is that we can never obtain the antenna pattern functions themselves, but only the ratios $F^i_{+,\times}/d_L$, which are degenerate with the luminosity distance.
Following this line of reasoning, we consider instead two ratios of angular functions: the relative channel power $Q_C$ and the relative polarization power $Q_P$.

\subsubsection{Relative channel power}

We start by defining the \emph{relative channel power} $Q_C$ between channels I and II:
\be\label{eq:Q1def}
Q_C=\frac{(F^{\rm I}_+)^2+(F^{\rm I}_\times)^2}{(F^{\rm II}_+)^2+(F^{\rm II}_\times)^2}\,.
\ee

\begin{figure}[t]
  \includegraphics[width=\columnwidth]{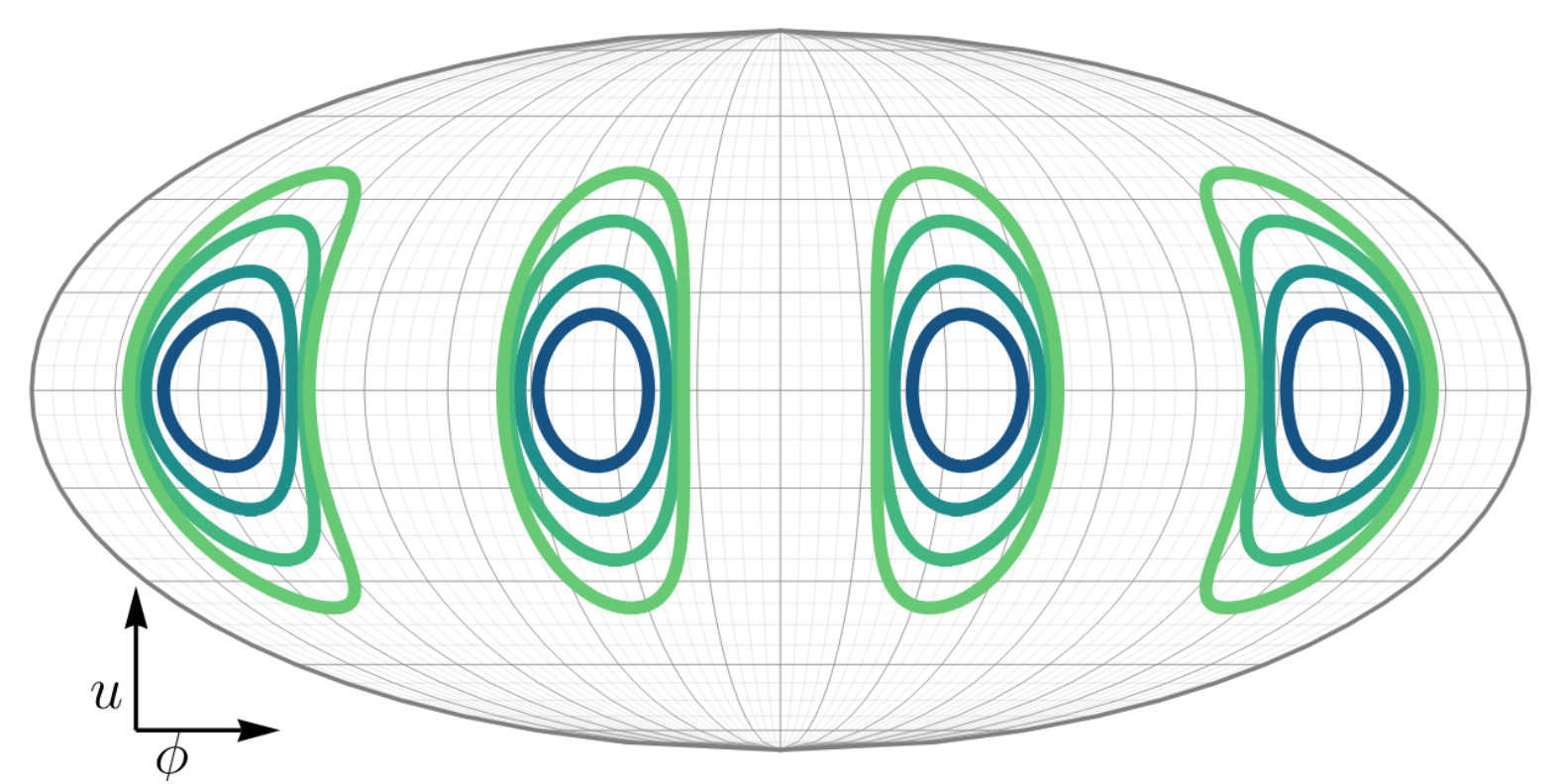}  
  \caption{Constant-$Q_C$ contours (Eq.~\ref{eq:ringEquation}) for $Q_C=0.25$ (innermost, dark blue contour), $0.5,\, 0.75$, and $0.9$ (outermost, light green contour).} 
\label{fig:localizationRingsQC}
\end{figure}

This combination has some interesting properties. First of all, the numerator and the denominator (which can be thought of as the \emph{antenna power} of each channel, or detector) are independent of the polarization angle $\psi$, and they are given by simple functions of $u=\cos\theta$ and $\phi$:
\be
(F^{ i}_+)^2+(F^{ i}_\times)^2=\frac{1}{8}\left[1+6 u^2+u^4\pm\left(u^2-1\right)^2 \cos (4 \phi)\right]\,,
\ee
where the plus sign corresponds to the first channel ($i={\rm I}$), while the minus sign corresponds to the second channel ($i={\rm II}$). Because of this property, constant-$Q_C$ contours in the sky can be found from the analytic relation
\be\label{eq:ringEquation}
\cos 4\phi= \frac{Q_C-1}{Q_C+1}\frac{( 1+6 u^2+u^4)}{ \left(u^2-1\right)^2}\,,
\ee
and they are shown in Fig.~\ref{fig:localizationRingsQC}.  The intersection of the constant-$Q_C$ contours of Fig.~\ref{fig:localizationRingsQC} with the localization contours of Fig.~\ref{fig:QAP} corresponds (in the absence of measurement errors) to a finite set of {\em points} in the sky.

The relative channel power $Q_C$ can be computed from the detector amplitudes as follows. One possibility is to solve Eq.~(\ref{eq:Adef}) to find $F_{+,\times}^{\rm I, II}/d_L$, and to use these quantities to compute $Q_C$. In alternative, we can use the relation
\be
(\Omega^{i}_{21})^2 - 4(\Omega^{i}_{22})^2  \propto (F^{ i}_+)^2+(F^{ i}_\times)^2\,,
\ee
to show that
\be\label{eq:Q1amp}
Q_C = Q_A \f{4 \hat{A}_{21}(q)^2-\left(\hat{\mathcal A}_{\rm I}\right)^2}{4 \hat{A}_{21}(q)^2- \left(\hat{\mathcal A}_{\rm II}\right)^2}\,,
\ee
where $\hat{A}_{21}(q)\equiv A_{21}(q)/A_{22}(q)$ is the relative mode amplitude, while $\hat{{\mathcal A}}_i\equiv {\mathcal A}^i_{21}/{\mathcal A}^i_{22}$ is the relative detector amplitude. 

\subsubsection{Relative polarization power}

A second useful combination is the \emph{relative polarization power}
\be\label{eq:Q2def}
Q_P=\frac{(F^{\rm I}_+)^2+(F^{\rm II}_+)^2}{(F^{\rm I}_\times)^2+(F^{\rm II}_\times)^2}\,.
\ee

This quantity is complementary to $Q_C$, in the following sense. First of all, the numerator and the denominator are now independent of the polarization angle $\phi$, and they are given by simple functions of $u=\cos\theta$ and $\psi$:
\be
(F^{\rm I}_p)^2+(F^{\rm II}_p)^2=\frac{1}{8}\left[1+6 u^2+u^4\pm\left(u^2-1\right)^2 \cos (4 \psi)\right]\,.
\ee
where the plus sign corresponds to the plus polarization, while the minus sign corresponds to the cross polarization.
By the same reasoning outlined above we find that
\be\label{eq:ringEquationQ2}
\cos 4\psi= \frac{Q_P-1}{Q_P+1}\frac{( 1+6 u^2+u^4)}{ \left(u^2-1\right)^2}\,,
\ee
and therefore constant-$Q_P$ contours are completely identical to those shown in Fig.~\ref{fig:localizationRingsQC} for $Q_C$.

By solving Eq.~(\ref{eq:Adef}) for $ F_{+,\times}^{\rm I, II}/d_L$ and using these quantities to calculate $Q_P$ we get
\be\label{eq:Q2amp}
Q_P = -\cos^2\iota \f{\hat{A}_{21}(q)^2-\left(\hat{{\mathcal A}}_{\rm (I + II)}/s^\times_{21}(\iota)\right)^2}{\hat{A}_{21}(q)^2-\left(\hat{{\mathcal A}}_{\rm (I + II)}/s^+_{21}(\iota)\right)^2}\,,
\ee
where we have defined
\be
\hat{{\mathcal A}}_{\rm (I + II)}^2 \equiv \f{Q_A (\hat{{\mathcal A}}_{\rm I})^2 + (\hat{{\mathcal A}}_{\rm II})^2}{Q_A+1}= 
\f{({\mathcal A}^{\rm I}_{21})^2 + ({\mathcal A}^{\rm II}_{21})^2}{({\mathcal A}^{\rm I}_{22})^2 + ({\mathcal A}^{\rm II}_{22})^2}
\ee
as well as 
\bea 
s_{21}^+(\iota) &\equiv& \f{Y_+^{21}(\iota)}{Y_+^{22}(\iota)} =\f{2 \sin\iota}{1+ \cos ^2\iota}\,, \nn\\
s_{21}^\times(\iota) &\equiv& \f{Y_\times^{21}(\iota)}{Y_\times^{22}(\iota)} =\sin\iota\,.
\eea

Constant-$Q_C$ and constant-$Q_P$ contours are both bounded in latitude: for example $-u_{m}(Q_P)<u<u_{m}(Q_P)$, where
\be
u_{m}(Q_P)=\sqrt{-1+\frac{2}{Q_P}-\frac{2 \sqrt{1-Q_P}}{Q_P}}\,.
\ee
An identical relation holds for $Q_C$.

The intersection of constant-$Q_P$ contours with the localization contours of Fig.~\ref{fig:QAP} also corresponds (at least in the absence of measurement errors) to a finite set of {\em points} in the sky. In both cases, when solving for sky position we inevitably end up with multiple solutions. The situation is not too dissimilar from sky localization with (say) three Earth-based interferometers: by using times of arrival for each two-detector combination we can identify a ring in the sky, and the intersection of two rings identifies {\em two} points in the sky.

Is there an optimal strategy to find ``the'' right solution in our case?
One possibility to further localize the signal is to use the time delay between different spacecraft. Time-delay contributions appear as higher-order corrections to the phase which depend on the projected arm lengths $L_{ij} = L\left(1-\hat{\textbf{n}}\cdot\hat{\textbf{r}}_{ij}\right)$, where $\hat{\textbf{r}}_{ij}$ denotes the unit separation vector between spacecraft $i$ and $j$, $L_{ij}$ is the corresponding arm length, and $\hat{\textbf{n}}$ is the unit vector pointing towards the source~\cite{Robson:2018jly}. These projected arm lengths can be related to the sky location, and therefore an accurate phase measurement could (in principle) give more insight on sky location. This method is more effective for high-frequency signals.

Ref.~\cite{Robson:2018jly} studied the localization of sine-Gaussian bursts by measuring time delays between different spacecraft, finding that bursts with short duration could be localized much better than bursts with longer duration due to a degeneracy between the central time of the burst wavelet and the sky localization: bursts with a longer duration yield poor constraints on the central time, and hence poor sky localization. Similar arguments should be applicable to ringdown signals. In the case of ringdown, the ``starting time'' $t_0$ in Eq.~(\ref{eq:hdef}) -- which is the analog of the central time in the burst analysis -- can be determined with good accuracy from relative phase calculations. In principle it should be possible to use higher-order phase corrections to improve the sky-localization procedure based on relative amplitudes that we described above .

\begin{figure*}[t]
\begin{tabular}{cc}
  \includegraphics[width=0.515\textwidth]{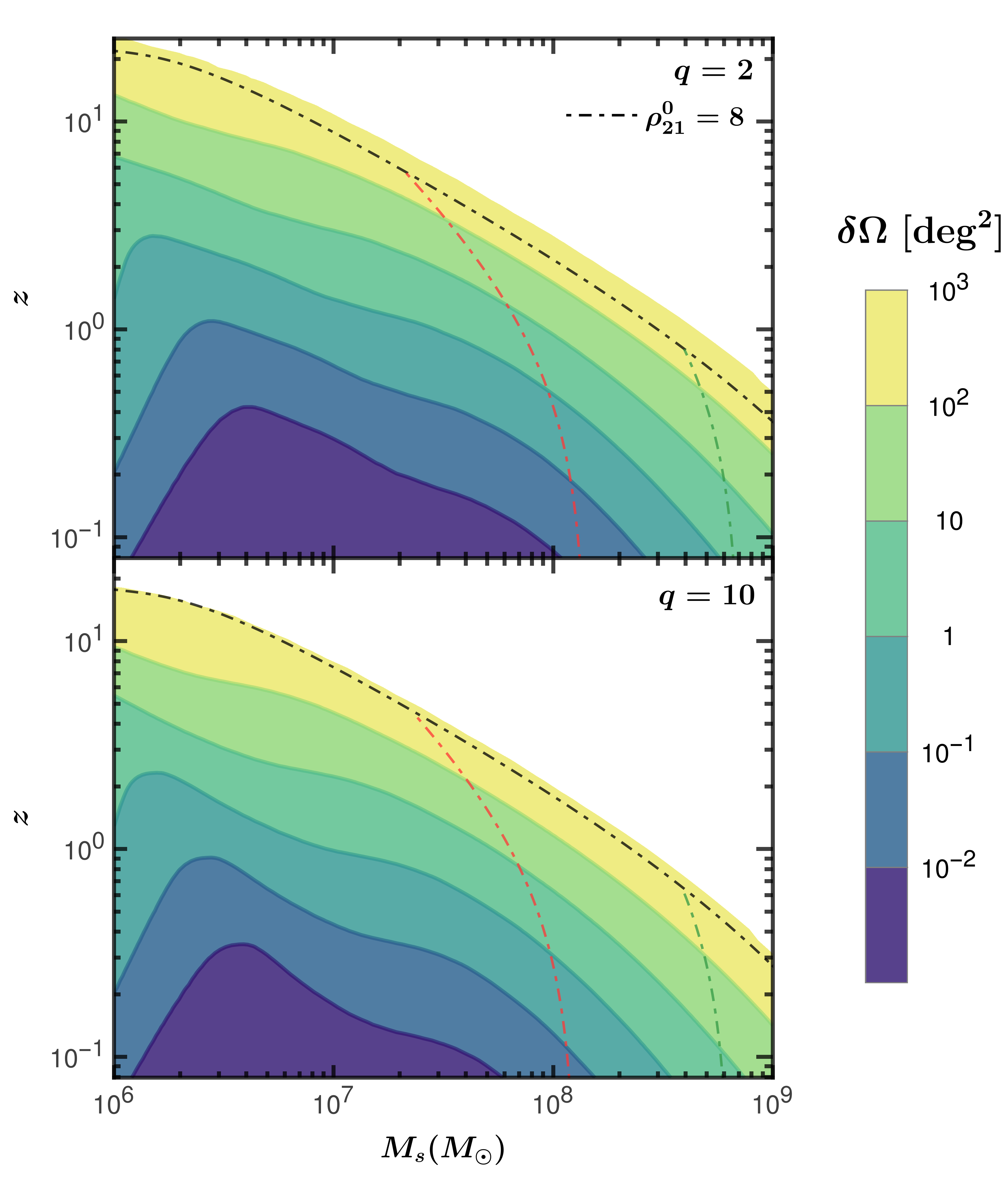}
  \includegraphics[width=0.484\textwidth]{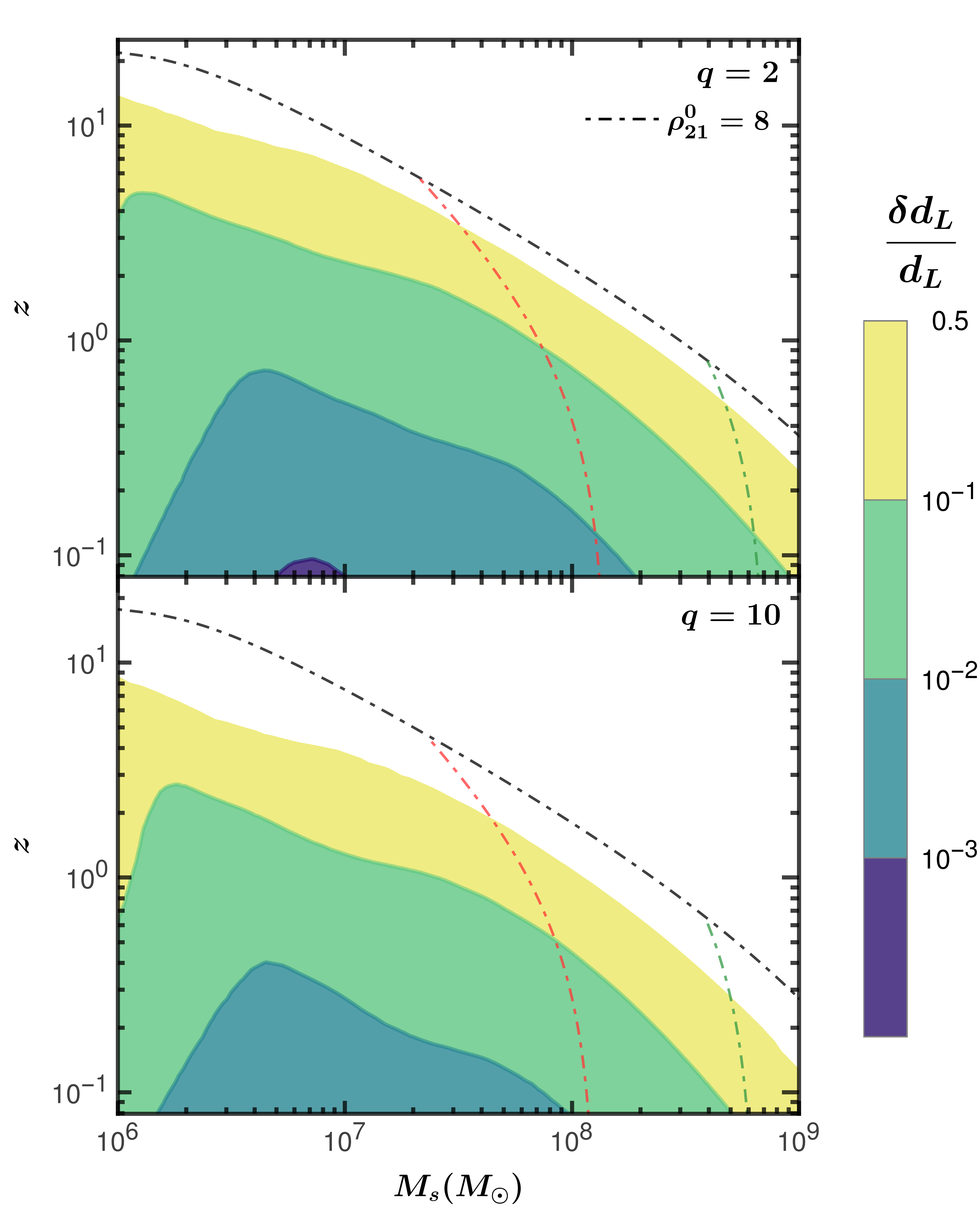}
\end{tabular}
  \caption{Median errors on sky localization (left) and luminosity distance (right) for binaries with $q=2$ (top) and $q=10$ (bottom). 
We also show the horizon of $(2,\,1)$ mode and redshifted-mass cutoff at $f_{\rm cut}=10^{-4}$~Hz (in red) $f_{\rm cut}=2\times10^{-5}$~Hz (in green). %
  } 
\label{fig:errOD}
\end{figure*}

\subsection{Errors}

Now that we have outlined the general procedure, let us turn to estimating the sky localization errors using error propagation.

We have two independent ways of calculating the source position and polarization: we can use either $(Q_A, Q_\Phi, Q_C)$ or $(Q_A, Q_\Phi, Q_P)$. The unknowns $\boldsymbol{\Theta_j}=\{\theta,\,\phi,\,\psi\}$ can be calculated from the three-vectors $\boldsymbol{Q_j}=\{Q_A,\,Q_\Phi,\,Q_j\}$ (where $j=C,\,P$). In turn, these three-vectors depend on the mass ratio $q$, the inclination $\iota$ and the detector amplitudes, which we will collectively denote as $\boldsymbol{X_{\Theta}}=\{q,\,\iota,\,\hat{\mathcal A}^{\rm I}_{21},\,\hat{\mathcal A}^{\rm II}_{21}\}$. Therefore we need a mapping between three sets of variables:
\be
\boldsymbol{X_{\Theta}} \to \boldsymbol{Q_j} \to \boldsymbol{\Theta_j}.
\ee
The covariance matrices for these sets of variables are related by Jacobian matrices as follows:
\be
\cov_j(\boldsymbol{\Theta})=\jacob{\boldsymbol{\Theta}}{\boldsymbol{Q_j}}\cdot\jacob{\boldsymbol{Q_j}}{\boldsymbol{X_{\Theta}}}\cdot \cov(\boldsymbol{X_{\Theta}}).\cdot\jacob{\boldsymbol{Q_j}}{\boldsymbol{X_{\Theta}}}^T \cdot \jacob{\boldsymbol{\Theta}}{\boldsymbol{Q_j}}^T
\ee
where a $T$ denotes the transpose.

We ignore errors on the amplitudes and phases of the $(2,\,2)$ mode, which are typically very small compared to the errors associated with $q$, $\iota$ or the $(2,\,1)$ amplitudes.
Furthermore we can neglect correlations between $\{q,\,\iota \}$ and the $(2,\,1)$ mode amplitudes, so the covariance matrix for $\boldsymbol{X_{\Theta}}$ is block-diagonal:
\be
\cov(\boldsymbol{X_{\Theta}})= \begin{pmatrix} 
\cov(q,\iota) & 0 & 0 \\
0 & 2 (\hat{{\mathcal A}_{21}^{\rm I}}/\rho^{\rm I}_{21})^2 & 0 \\
0 &  0 & 2 (\hat{{\mathcal A}_{21}^{\rm II}}/\rho^{\rm II}_{21})^2 
\end{pmatrix}
\ee

The Jacobian $\jacob{\boldsymbol{Q_j}}{\boldsymbol{X_{\Theta}}}$
can be calculated from Eqs.~(\ref{eq:Q1amp}) and (\ref{eq:Q2amp}), while the
Jacobian
$\jacob{\boldsymbol{\Theta}}{\boldsymbol{Q_j}}=\jacob{\boldsymbol{Q_j}}{\boldsymbol{\Theta}}^{-1}$ can be computed from Eqs.~(\ref{eq:QAdef}), (\ref{eq:QPdef}), (\ref{eq:Q1def}) and (\ref{eq:Q2def}).
 
It is possible to reduce the error by combining results from both $Q_C$ and $Q_P$:
\be
\cov(\boldsymbol{\Theta})^{-1}=\sum_{j}\cov_j(\boldsymbol{\Theta})^{-1}\,.
\ee

We define the sky-localization error as the determinant of the $(u,\phi)$-block of $\cov({\bf \Theta})$:
\be
\delta\Omega =\left\{{\rm det}\left[\cov (u,\phi)\right]\right\}^{1/2}\,.
\label{eq:dW}
\ee

In the left panel of Fig.~\ref{fig:errOD} we plot the median sky-localization errors for sources uniformly distributed over the sky. LISA can localize a  $M_s=10^6 M_\odot$ source with $q=2\,(10)$ within $100\,{\rm deg}^2$ up to redshift $z\approx 13\,(9.4)$. However sky localization relies on measurements of the $(2,\,1)$ mode, which has lower frequency than the $(2,\,2)$ mode (for fixed $M_s$) and gets out of band earlier as we increase the mass. Therefore sky-localization accuracy suffers at high masses: for example, we can localize a  $M_s=10^8 M_\odot$ source with $q=2\,(10)$ within $100\,{\rm deg}^2$ only up to redshift $z\approx1.7\,(1.2)$. It may be possible to localize such high-mass sources using the time evolution of the antenna pattern. This is because, as we show in Appendix~\ref{sec:evolutionLocalization}, the time-evolution of the amplitude is known much better than the $(2,\,1)$ amplitude for binaries with $M_s\gtrsim 5\times10^8 M_\odot$. In these cases, we may expect the errors to be significantly smaller.

In Fig.~\ref{fig:errOD} we show the ``reduced'' error obtained by combining both $Q_C$ and $Q_P$, but using $Q_P$ alone gives better sky-localization accuracy than using $Q_C$ alone for most sources (approximately $77\%$ of the sky).  This can be understood as follows. The relative channel power $Q_C$ [Eq.~(\ref{eq:Q1def})] and the amplitude ratio $Q_A$ [Eq.~(\ref{eq:QAdef})] differ only by factors of $s_{22}$ multiplying $F^i_{\times}$ in the numerator and in the denominator. From Fig.~\ref{fig:slm} we see that $s_{22}\simeq 1$ unless $\iota\gtrsim 90\deg$ (i.e., unless the binary inclination is close to edge-on).  We conclude that $Q_C \simeq Q_A$ in a large portion of the parameter space, and using $Q_C$ does not necessarily lead to new information.

Note that we chose to consider $Q_C$ and $Q_P$ mainly because they are easy to understand and manipulate, but in data analysis applications other combinations may be easier to measure, and the particular combination that leads to the smallest errors will in general depend on the source position and orientation. Some examples of combinations that could be considered include $F_+^{\rm I}/F_\times^{\rm I}$, $F_+^{\rm I}/F_+^{\rm II}$ , $F_+^{\rm I}/F_\times^{\rm II}$, etcetera.

\begin{table*}
  \caption{Redshifts at which various median errors are equal to the values indicated in the top row, for selected values of the remnant's source-frame mass $M_s$.}
\label{tab:zaterror}
\setlength\tabcolsep{6 pt}
\begin{tabular}{LCCCCCC}
\hline
\\
M_{\rm s}(M_\odot) & \f{\delta M}{M} =0.1\,(10^{-2}) & \delta a =0.1\,(10^{-2})& \f{\delta q}{q} =0.1\,(10^{-2}) & \delta\iota =10\degree\,(1\degree)  &{\delta\Omega}=10\ {\rm deg}^2\,(1\ {\rm deg}^2) & \f{\delta d_L}{d_L} =0.1\,(10^{-2})\\ 
\hline
\multicolumn{7}{C}{q=2}\\
\hline
 10^6 & {50\,(12)} & {40\,(9.8)} & {16\,(0.3)} & {18\,(0.3)} & \
{6.8\,(1.4)} & {3.6\,(0.05)} \\
 10^7 & {16\,(5.5)} & {14\,(4.5)} & {7.3\,(1.5)} & {8.1\,(1.7)} & \
{3.\,(1.4)} & {2.3\,(0.5)} \\
 10^8 & {4.1\,(1.5)} & {3.5\,(1.2)} & {2.1\,(0.7)} & {2.4\,(0.8)} & \
{0.9\,(0.5)} & {0.7\,(0.2)} \\
 10^9 & {0.9\,(0.2)} & {0.7\,(0.2)} & {0.4\,(0.06)} & {0.4\,(0.08)} & \
{0.1\,(0.04)} & {0.07\,(0.01)} \\
\hline
\multicolumn{7}{C}{q=10}\\
\hline
 10^6 & {22\,(5.9)} & {14\,(2.6)} & {0.7\,(0.04)} & {11\,(0.4)} & \
{5.5\,(1.3)} & {0.6\,(0.04)} \\
 10^7 & {9.7\,(2.5)} & {6.1\,(1.4)} & {1.3\,(0.3)} & {4.8\,(1.0)} & \
{2.2\,(1.0)} & {1.3\,(0.3)} \\
 10^8 & {2.6\,(0.8)} & {1.8\,(0.5)} & {0.5\,(0.09)} & {1.4\,(0.4)} & \
{0.6\,(0.3)} & {0.4\,(0.08)} \\
 10^9 & {0.5\,(0.09)} & {0.3\,(0.04)} & {0.04\,(0.005)} & {0.2\,(0.03)} \
& {0.05\,(0.02)} & {0.03\,(0.003)} \\
\hline
\end{tabular}
\end{table*}

\section{Luminosity distance}
\label{sec:lumdistance}

The strategy for sky localization in Sec.~\ref{sec:skyloc} was to determine the ratios $F^{\rm I,II}_{+,\times}/d_L$ between the antenna pattern functions and the luminosity distance. The antenna pattern functions depend on the angles $(\theta,\,\phi,\,\psi)$, so we can (at least in principle) determine these angles from a knowledge of $F^{\rm I,II}_{+,\times}/d_L$. At this point it would be straightforward to compute $d_L$.

A simple way to determine $d_L$ is to use the fact that the ``total'' antenna power depends only on $u=\cos\theta$:
\be
{\mathcal P}(u)=\sum_{i}\left[(F^i_+)^2+(F^i_\times)^2\right]=\frac{1}{4} \left(1+6 u^2+u^4\right)\,.
\ee
Then we can compute the distance in terms of the detector amplitudes of the $(2,\,2)$ and $(2,\,1)$ modes as follows:
\be
\frac{{\mathcal P}(u)}{d_L^2}=\frac{\zeta (\iota )}{M^2}\sum_{i={\rm I, II}}\left(\frac{\mathcal{A}_{22}^i}{A_{22}(q)}\right)^2\left[4-\left(\frac{\hat{\mathcal{A}}_{21}^i}{\hat{A}_{21}(q)}\right)^2\right]\,,
\ee
where
\be
\zeta(\iota)=\frac{4 \pi}{5}\frac{ \sec ^2\iota }{3+\cos ^2\iota }\,.
\ee
Next we estimate errors on the luminosity distance by error propagation. The unknown luminosity distance $d_L$ can be computed in the ``basis'' $\boldsymbol{X_d^j}=\{u_j,\,q,\,\iota,\,\hat{\mathcal A}^{\rm I}_{21},\,\hat{\mathcal A}^{\rm II}_{21}\}$, where $u_j$ is the colatitude calculated using $\boldsymbol{Q_j}=(Q_C,\,Q_P)$.
We will ignore once again the errors on the amplitude and phase of the $(2,\,2)$ mode, which are much smaller than the errors associated with $q$, $\iota$ or the $(2,\,1)$ amplitudes. Then we have
\be
(\delta d_L)^{-1}=\sum_i\left(\jacob{ d_L }{\boldsymbol{X_d^j}}\cdot \cov(\boldsymbol{X_d^j})\cdot\jacob{d_L}{\boldsymbol{X}_d^j}^T\right)^{-1}\,.
\ee
Since correlations between $\{q,\,\iota\}$ and the $(2,\,1)$ mode amplitudes are negligible
and we are ignoring the errors associated with the $(2,\,2)$ mode, the covariance matrix for $\boldsymbol{X_d^j}$ is simply
\be
\cov(\boldsymbol{X_d^j})= \begin{pmatrix} 
\delta u_j & \cov(u, \boldsymbol{X_\Theta}) \\
\cov(u, \boldsymbol{X_\Theta})^T &  \cov(\boldsymbol{X_\Theta})\,,
\end{pmatrix}
\ee
where $\cov(u, \boldsymbol{X_\Theta})$ reads
\be
\cov(u, \boldsymbol{X_\Theta})=\jacob{u}{\boldsymbol{Q_j}}\cdot\jacob{\boldsymbol{Q_j}}{\boldsymbol{X_\Theta}}\cdot \cov(\boldsymbol{X_\Theta})\,.
\ee

Even if we have no sky localization information, we can still compute an ``effective distance'' $d_\star$ defined as follows:
\be
d_\star=\f{d_L}{\sqrt{4\sum_{i}\left[(F^i_+)^2+(F^i_\times)^2 \right]}}\,.
\ee
This quantity is very similar to the ``effective distance'' for LIGO-like Earth-based detectors, which is degenerate with the inclination angle $\iota$~\cite{Chen:2017wpg}. %

Even in the worst-case scenario where $u$ is completely unconstrained, the allowed range for $d_\star$ is relatively limited: $d_\star \le d_L\le 2\sqrt{2} d_\star$. However in most cases the $(2,\,2)$ mode is dominant, so $Q_A$ and $Q_\Phi$ can be determined very accurately. These quantities alone cannot determine the sky location, but they can be used to set bounds on $u$ which can be very narrow (especially when the inclination is not close to edge-on): see for example the $\iota=45\degree$ case in the top panel of Fig.~\ref{fig:QAP}, for which $0.47 <\lvert u \rvert< 0.58$, or $1.54\,d_\star<d_L <1.77\,d_\star$.

In the right panel of Fig.~\ref{fig:errOD} we plot the median luminosity distance errors for sources uniformly distributed and oriented over the sky. The top panels show that for a binary with $q=2$, LISA could measure $d_L$ with an accuracy of $10\%$ up to redshift $z=3.6$ for BHs of mass $10^6 M_\odot$. In the bottom panels we consider a binary with $q=10$, and we show that LISA could measure $d_L$ with better than $10\%$ accuracy out to $z=0.6$ for $M_s=10^6 M_\odot$.

Table~\ref{tab:zaterror} summarizes LISA's parameter estimation capabilities by listing the redshift out to which various median errors are equal to specific thresholds (indicated in the top row) for selected values of the remnant's source-frame mass $M_s$.

\begin{figure}[t]
\begin{tabular}{c}
  \includegraphics[width=0.95\columnwidth]{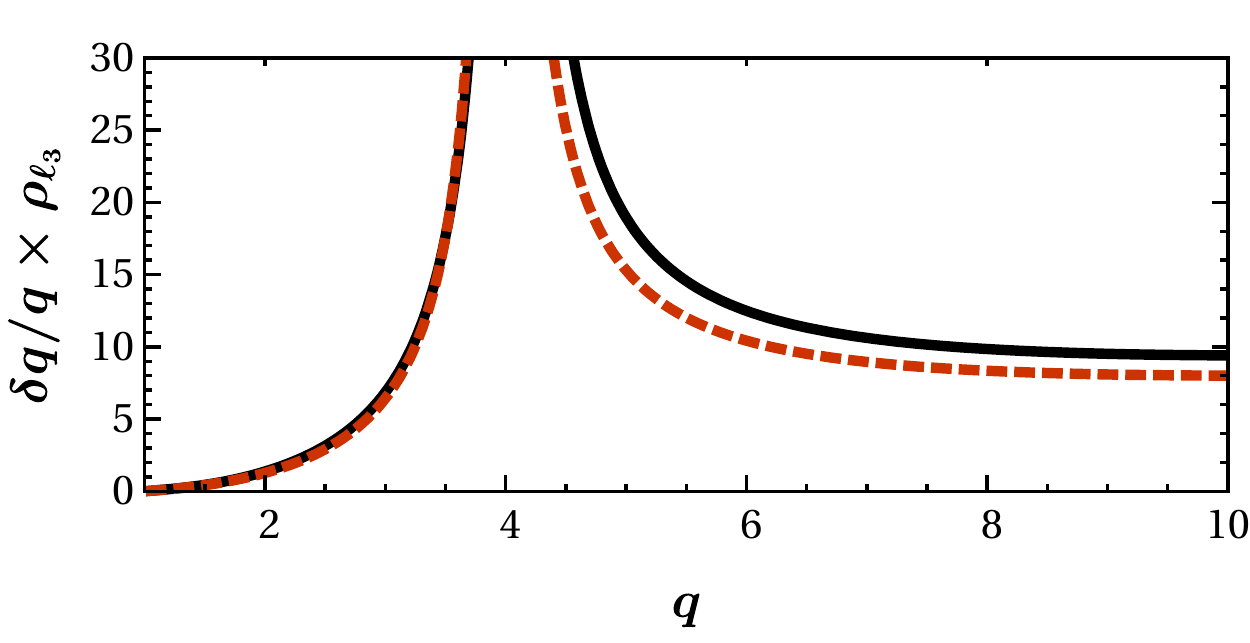}\\
  \includegraphics[width=0.95\columnwidth]{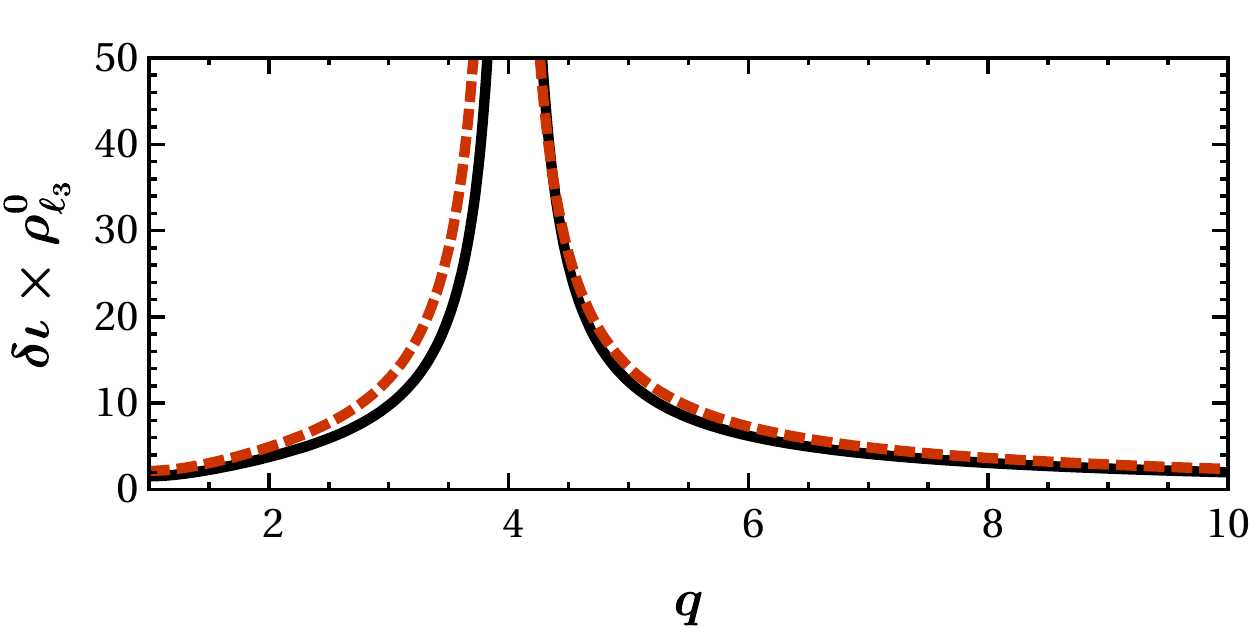}\\  
  \includegraphics[width=0.95\columnwidth]{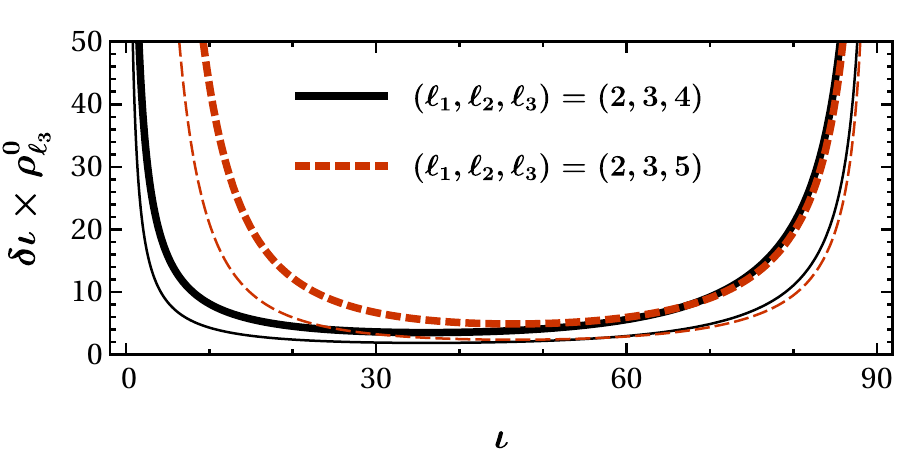}
 \end{tabular}
  \caption{Top panel: Relative error $\delta q/q$ on the mass ratio, scaled by the SNR $\rho_{\ell_3}$ of the third (least dominant) mode used in the analysis.  Middle panel: inclination error $\delta \iota$ scaled by the optimal SNR $\rho^0_{\ell_3}$ of the third (least dominant) mode used in the analysis as a function of $q$, for $\iota=45\degree$. Bottom panel: inclination error $\delta \iota$ scaled by the optimal SNR $\rho^0_{\ell_3}$ as a function of $\iota$ for $q=2$ (thick lines) and $q=10$ (thin lines).} 
\label{fig:errQIvsQI}
\end{figure}

\section{Error dependence on mass ratio, inclination and sky position}
\label{sec:qIotaSkyDep}

So far we have mostly estimated errors for specific binary systems. We now wish to explore more systematically the dependence of the errors on the mass ratio $q$, the inclination $\iota$, and the sky position of the source.

\subsection{Mass-ratio and inclination dependence}

Let us start by exploring the $q$-dependence of the errors. We consider a three-mode combination as in Eq.~(\ref{eq:ampRatio2}) and assume that $\ell_3$ is the least dominant mode. If we ignore the errors on the dominant modes and we also ignore correlations, we can show from Eq.~(\ref{eq:ampRatio2}) that the error on $q$ can be written as
\be\label{eq:dq}
\delta q=\frac{\sqrt{2}}{\rho_{\ell_3}} \frac{ G_{\ell_1 \ell_2 \ell_3}}{G_{\ell_1 \ell_2 \ell_3}'}\,,
\ee
where a prime denotes a derivative with respect to $q$. Recall that according to Eq.~(\ref{eq:rho0}) the SNR in a given mode can be factored as $\rho_{\ell m}= \rho_{\ell m}^0 \times w_{\ell m}$, where $\rho_{\ell m}^0$ is the SNR for an optimally oriented binary, and $w_{\ell m}(\iota,\theta,\phi,\psi)$ is a position, orientation and polarization-dependent ``projection factor'' such that $0\le w_{\ell m}\le 1$ (see e.g.~\cite{Dominik:2014yma}). 

For most binaries, the two strongest modes correspond to $\ell_1=\ell=m=2$ and $\ell_2=\ell=m=3$ (see e.g.~\cite{Baibhav:2018rfk}). In Fig.~\ref{fig:errQIvsQI} we plot the errors on various quantities assuming that either $\ell_3=4$ or $\ell_3=5$. In both cases the fractional error $\delta q/q$ diverges at $q\approx4$ because $G_{\ell_1 \ell_2 \ell_3}'=0$ there (cf. Fig.~\ref{fig:G}) and it saturates at large $q$, approaching the limit
\bea
\f{\delta q}{q}\approx \f{9.4}{\rho_{44}}\,,\quad \f{\delta q}{q}\approx \f{8.0}{\rho_{55}}\,.
\eea

For the inclination we find
\bea\label{eq:deltaidep}
\delta \iota  &=&\frac{\tan  \iota }{\ell_2-\ell_1} \delta q \left| \frac{H_{\ell_1 \ell_2}'}{H_{\ell_1 \ell_2} }\right|\,\\
&=&\frac{1}{\rho_{\ell_3}}\frac{\tan  \iota }{\ell_2-\ell_1} \left| \frac{H_{\ell_1 \ell_2}'}{H_{\ell_1 \ell_2} }  \frac{ G_{\ell_1 \ell_2 \ell_3}}{ G_{\ell_1 \ell_2 \ell_3}'}\right|\,,
\eea
and the error diverges at $q\approx4$ for the same reason.

\begin{figure}[t]
\begin{tabular}{c}
  \includegraphics[width=0.95\columnwidth]{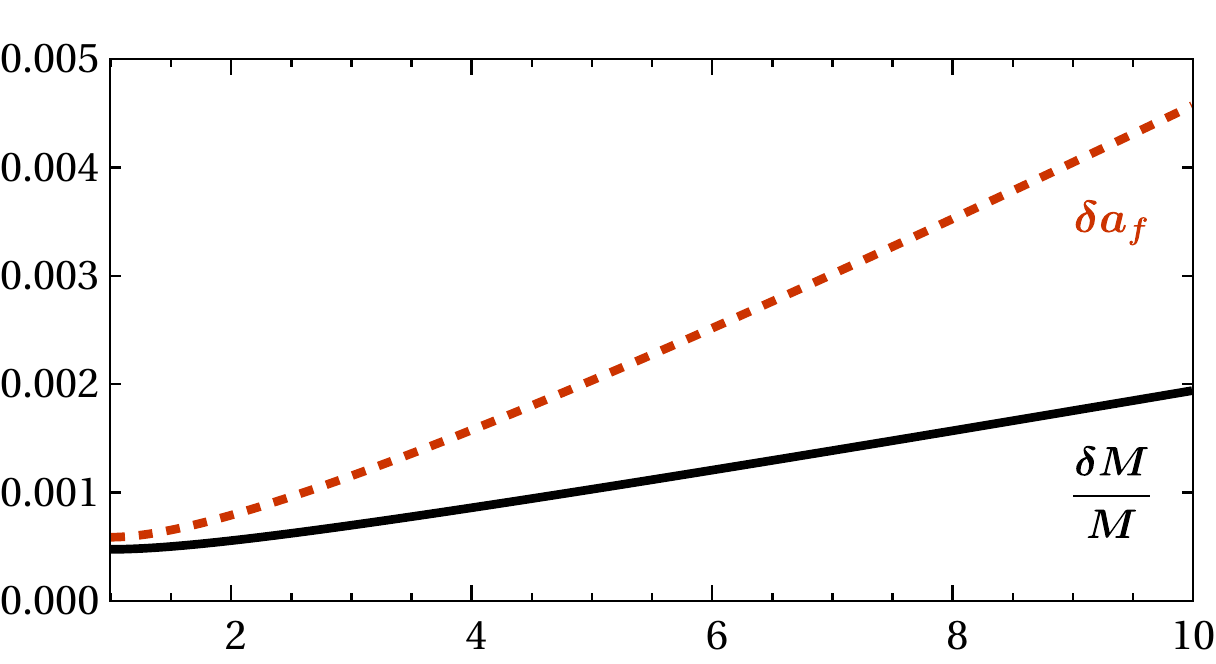}\\
  \includegraphics[width=0.95\columnwidth]{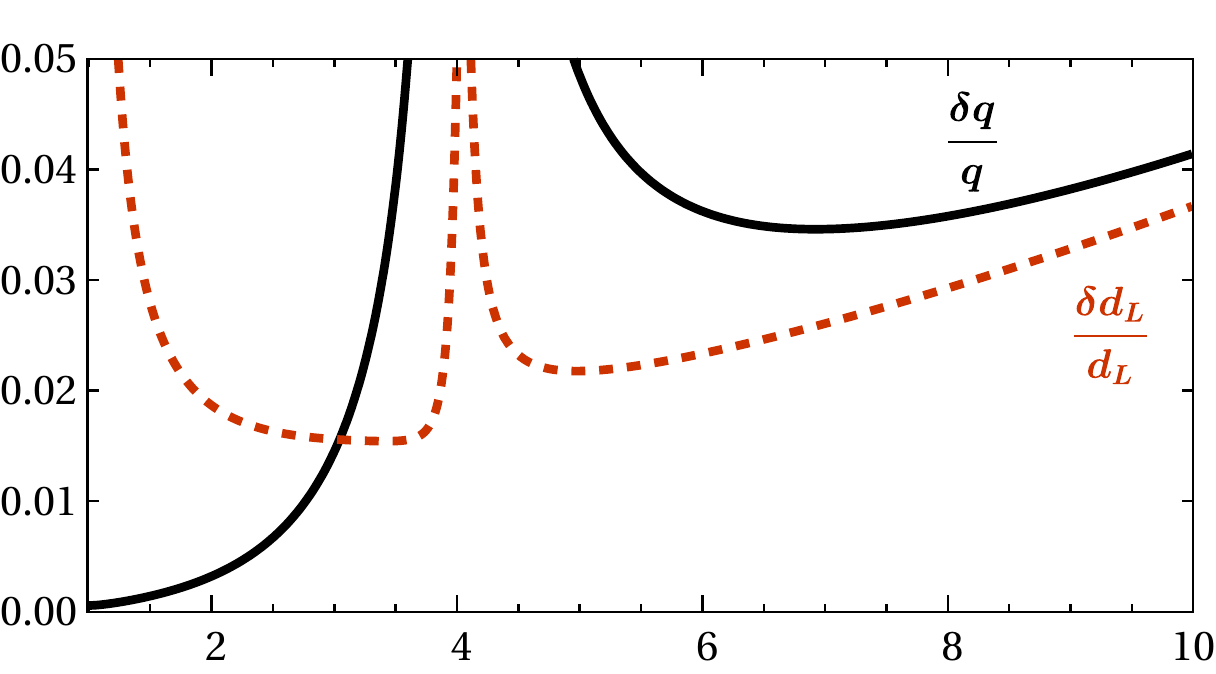}\\
  \includegraphics[width=0.95\columnwidth]{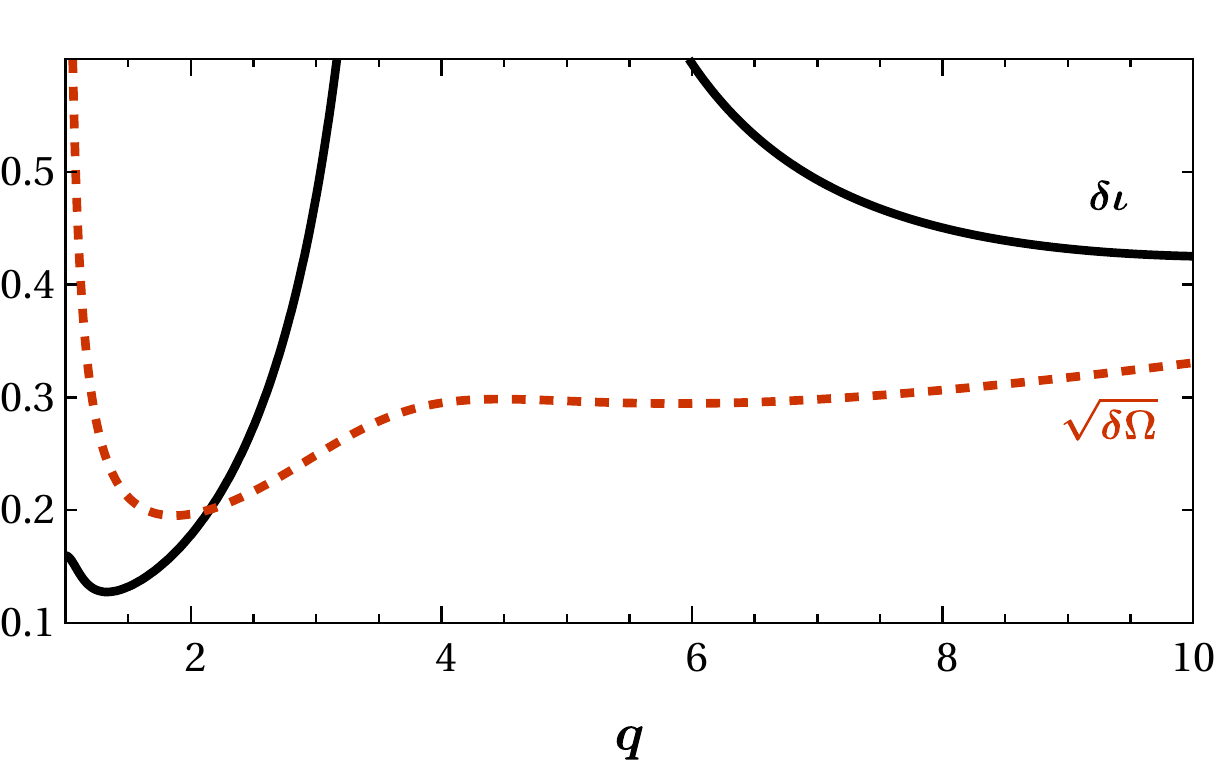}
  \end{tabular}
  \caption{Mass-ratio dependence of various errors for a $M_s=10^7 M_\odot$ remnant at $z=1$ with $(\iota,\,u,\,\phi,\,\psi)=(45\degree,\,0.5,\,30\degree,\,60\degree)$. %
  } 
\label{fig:Qdep}
\end{figure}

Finding analytical scalings for the errors on $d_L$ and $\Omega$ is not as simple, mainly because the sky-position dependent terms are complex and we have to ``change basis'' twice, as explained above. In Fig.~\ref{fig:Qdep} we consider for definiteness a $M_s=10^7 M_\odot$ remnant at $z=1$ with $(\iota,\,u,\,\phi,\,\psi)=(45\degree,\,0.5,\,30\degree,\,60\degree)$, and we plot the $q$-dependence of various errors. Mass and spin errors depend on the remnant properties, which in turn depend on $q$.  As expected, $\delta q/q$ and $\delta \iota$ diverge close to $q=4$, and the errors are typically smallest for small values of $q$. Interestingly, the sky-localization errors have a weaker dependence on $q$ and they do not diverge at $q=4$, but they do diverge for nearly equal-mass systems ($q\to 1$). Distance errors diverge at both $q\simeq 1$ and $q\simeq 4$. %

\begin{figure}[htp]
\begin{tabular}{c}
  \includegraphics[width=0.95\columnwidth]{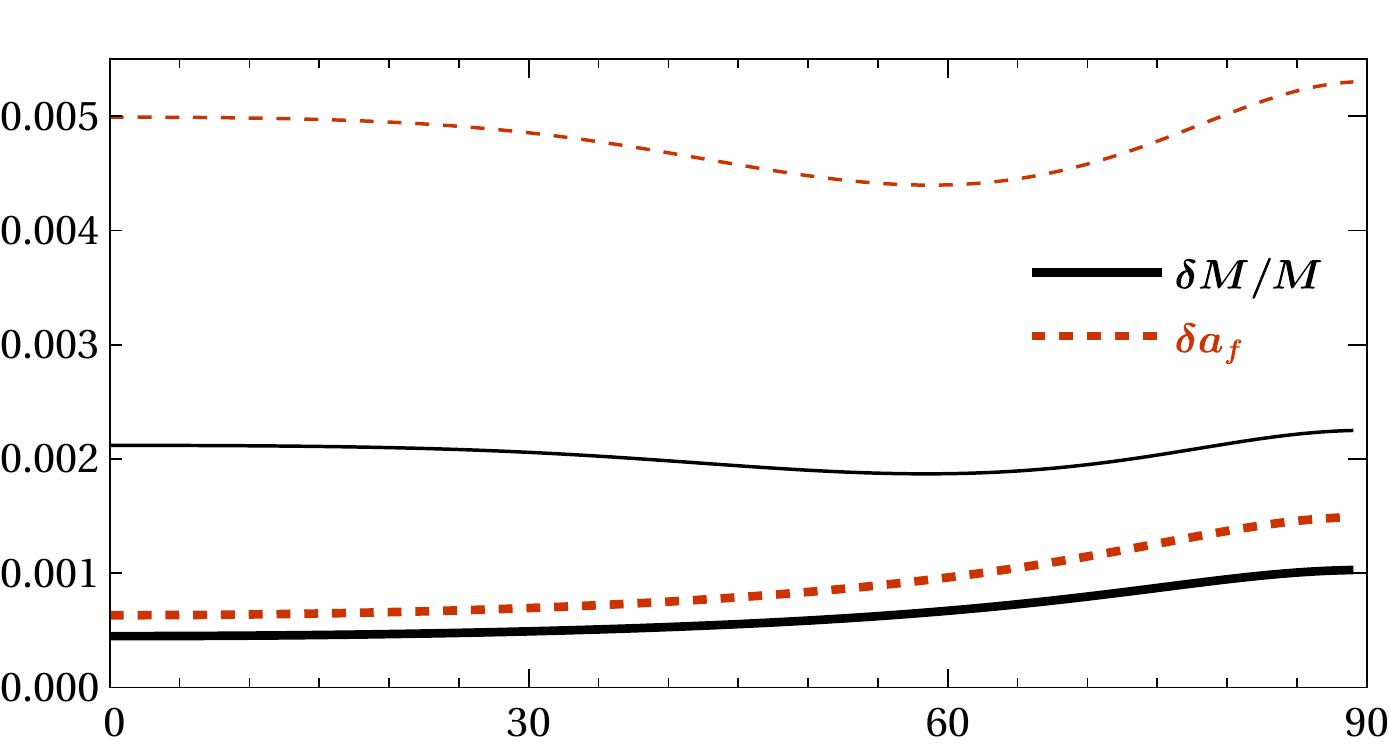}\\
  \includegraphics[width=0.95\columnwidth]{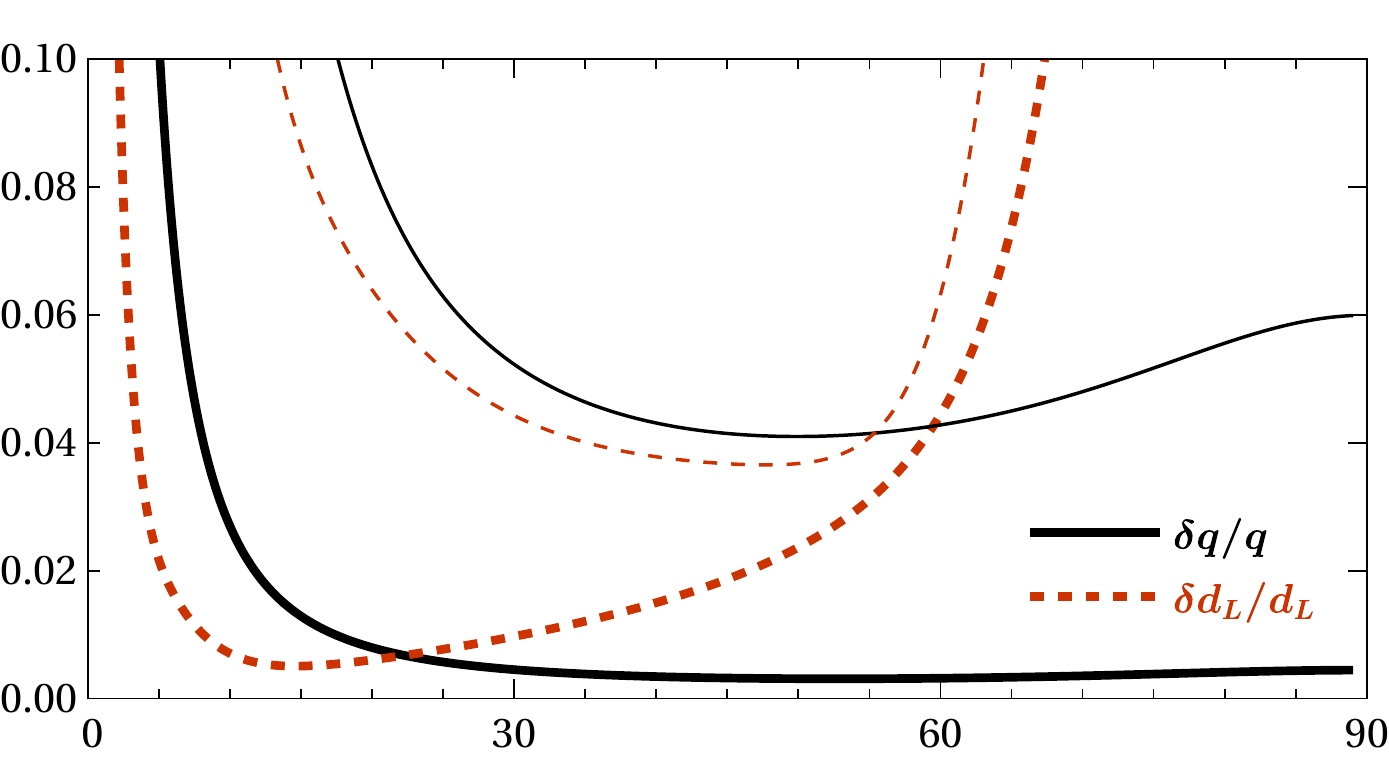}\\
  \includegraphics[width=0.95\columnwidth]{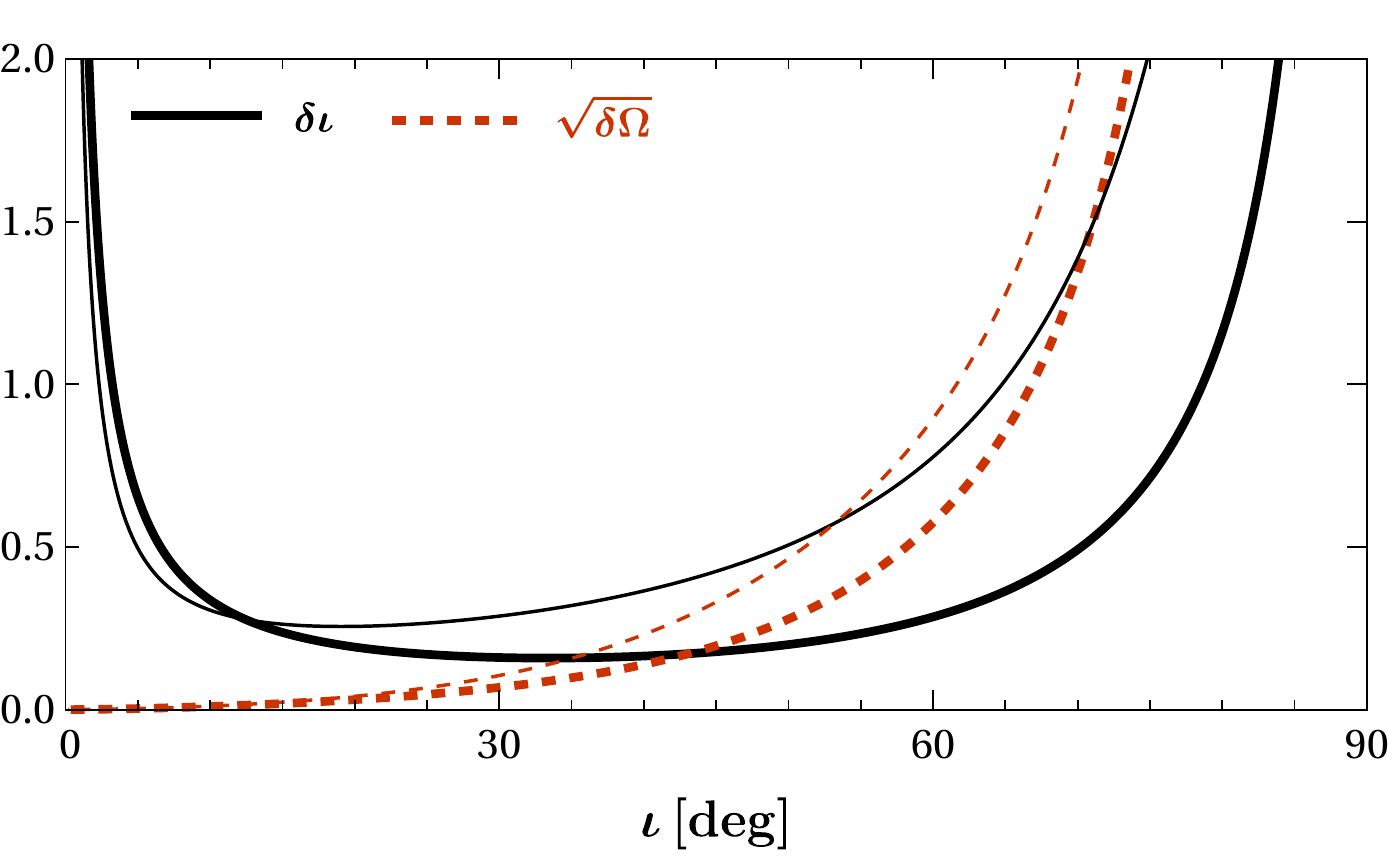}
  \end{tabular}
  \caption{Inclination dependence of various errors for a $M_s=10^7 M_\odot$ remnant at $z=1$ with $(u,\,\phi,\,\psi)=(0.5,\,30\degree,\,60\degree)$ and $q=2$ (thick lines) or $q=10$ (thin lines).}
\label{fig:Idep}
\end{figure}

Equation~(\ref{eq:dq}) for $\delta q$ depends on the inclination $\iota$ only through $\rho_{\ell_3}$.  To single out the $\iota$ dependence, we average the projection factor $w_{\ell m}(\iota,\theta,\phi,\psi)$ over the remaining angles ($\theta$, $\phi$ and $\psi$) with the result
\be
\delta q \propto \frac{1}{\bar{w}_{\ell_3\ell_3}(\iota)}\propto \frac{1}{(\sin\iota)^{\ell_3-2} \sqrt{1+6 \cos ^2\iota+\cos ^4\iota}}\,,
\ee
which diverges for face-on binaries. By proceeding in a similar way we find that, upon angle-averaging, $\delta \iota$ in Eq.~(\ref{eq:deltaidep}) reduces to
\bea
\delta \iota & \propto& \frac{\tan\iota }{\bar{w}_{\ell_3\ell_3}(\iota)}\nn\\
& \propto&  \frac{1}{(\sin\iota)^{\ell_3-3} \cos\iota\sqrt{1+6 \cos ^2\iota+\cos ^4\iota}}\,,
\eea
which diverges for {\em both} face-on and edge-on binaries (as shown in the bottom panel of Fig.~\ref{fig:errQIvsQI}). This can be understood as follows. The amplitude of $\ell=m$ modes is proportional to $\sin\iota^{\ell-2}$, so the amplitude of higher harmonics is very low for face-on binaries. On the other hand, for edge-on binaries $\sin\iota^{\ell-2}$ is flat, and measuring $\iota$ is hard.

Let us now look at the $\iota$ dependence of various errors. In Fig.~\ref{fig:Idep} (which is similar to Fig.~\ref{fig:Qdep}) we consider for definiteness a $M_s=10^7 M_\odot$ remnant at $z=1$ with $(u,\,\phi,\,\psi)=(0.5,\,30\degree,\,60\degree)$, and we plot the $i$-dependence of various errors for two selected values of the mass ratio ($q=2,\,10$).

Some remarks are in order. Spin and mass errors ($\delta a_f$ and $\delta M/M$) depend on $\iota$ only through the joint SNR $(\rho_\lm^{\rm I})^2+(\rho_\lm^{\rm II})^2 \propto (\Omega_\lm^{\rm I})^2+(\Omega_\lm^{\rm II})^2$. For moderate mass ratios ($q=2$) the $(2,\,2)$ mode is dominant, and higher harmonics do not contribute much to the measurement of $a_f$ and $M$.
For the $(2,\,2)$ mode, $(\Omega_{22}^{\rm I})^2+(\Omega_{22}^{\rm II})^2$ decreases with $\iota$, leading to smaller SNRs and larger errors for edge-on binaries.
The situation is different for larger mass ratios ($q=10$): higher harmonics are more prominent, and their contribution to the error budget is comparable to the $(2,\,2)$ mode (cf. Fig.~\ref{fig:errRemnantz1}). The higher harmonics vanish when the binary is face-on -- i.e. when most of the SNR comes from $(2,\,2)$ mode -- and have maxima when $0<\iota<\pi/2$, unlike the $(2,\,2)$, which decreases monotonically with $\iota$. As a result, $\delta a_f$ and $\delta M/M$ have a minimum when $q=2$.

We can also use Fig.~\ref{fig:Idep} to better understand Fig.~\ref{fig:Qdep}, in which we had fixed $\iota=45\degree$. For example, from the bottom panel of Fig.~\ref{fig:Idep}  we see that face-on binaries ($\iota\simeq 0\degree$) have similar inclination errors for $q=2$ and $q=10$, while for edge-on binaries ($\iota\simeq 90\degree$) $\delta\iota$ is  larger for $q=10$ than for $q=2$. In Fig.~\ref{fig:Qdep}, the mass ratio dependence would have been milder (stronger) had we considered $\iota\simeq 0\degree$ ($\iota\simeq 90\degree$) rather than $\iota=45\degree$. Inclination has a much milder effect on sky localization errors, whether $q=2$ or $q=10$.

\begin{figure}[t]
\begin{tabular}{c}
  \includegraphics[width=\columnwidth]{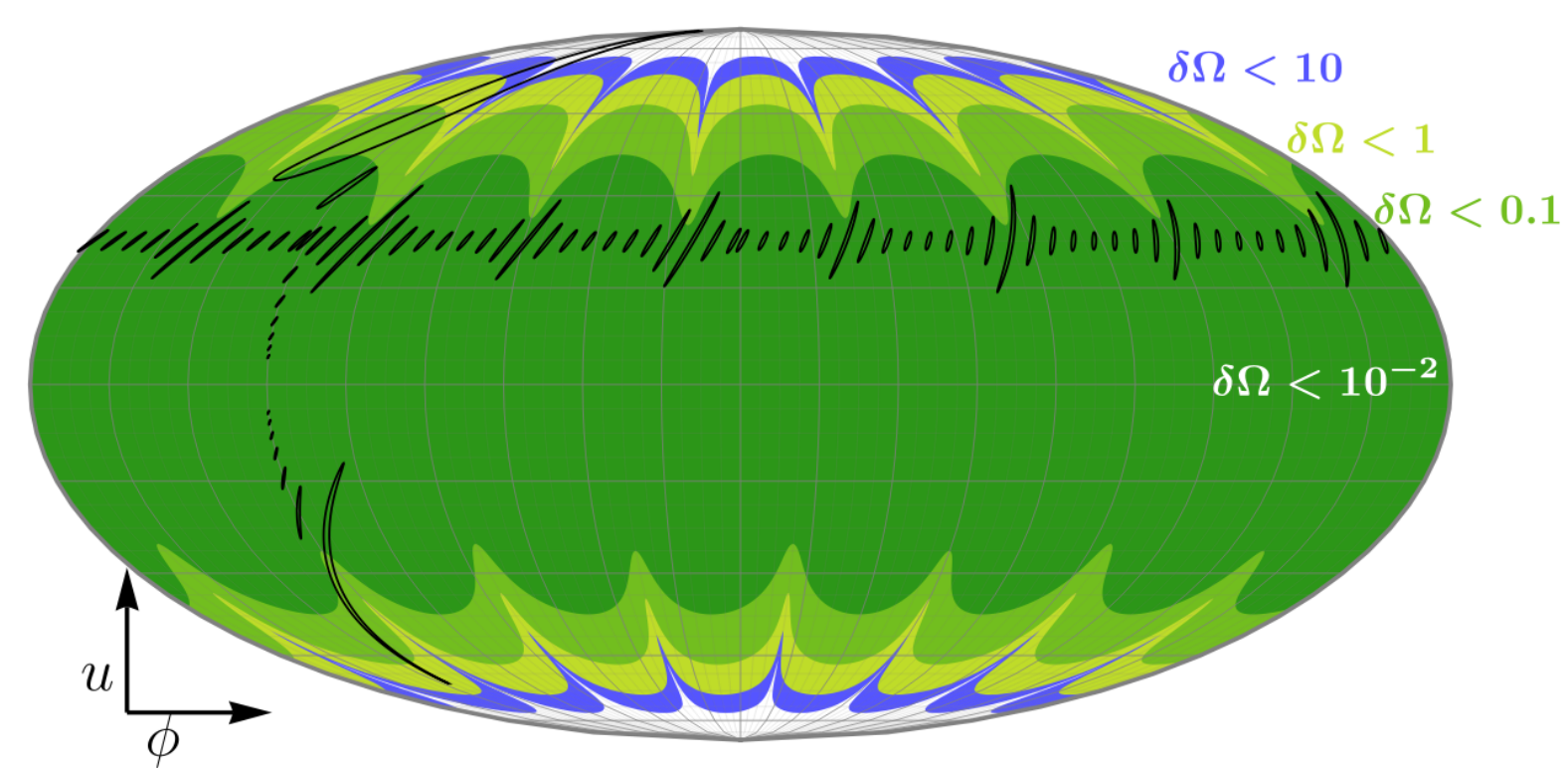}\\
   \includegraphics[width=0.9\columnwidth]{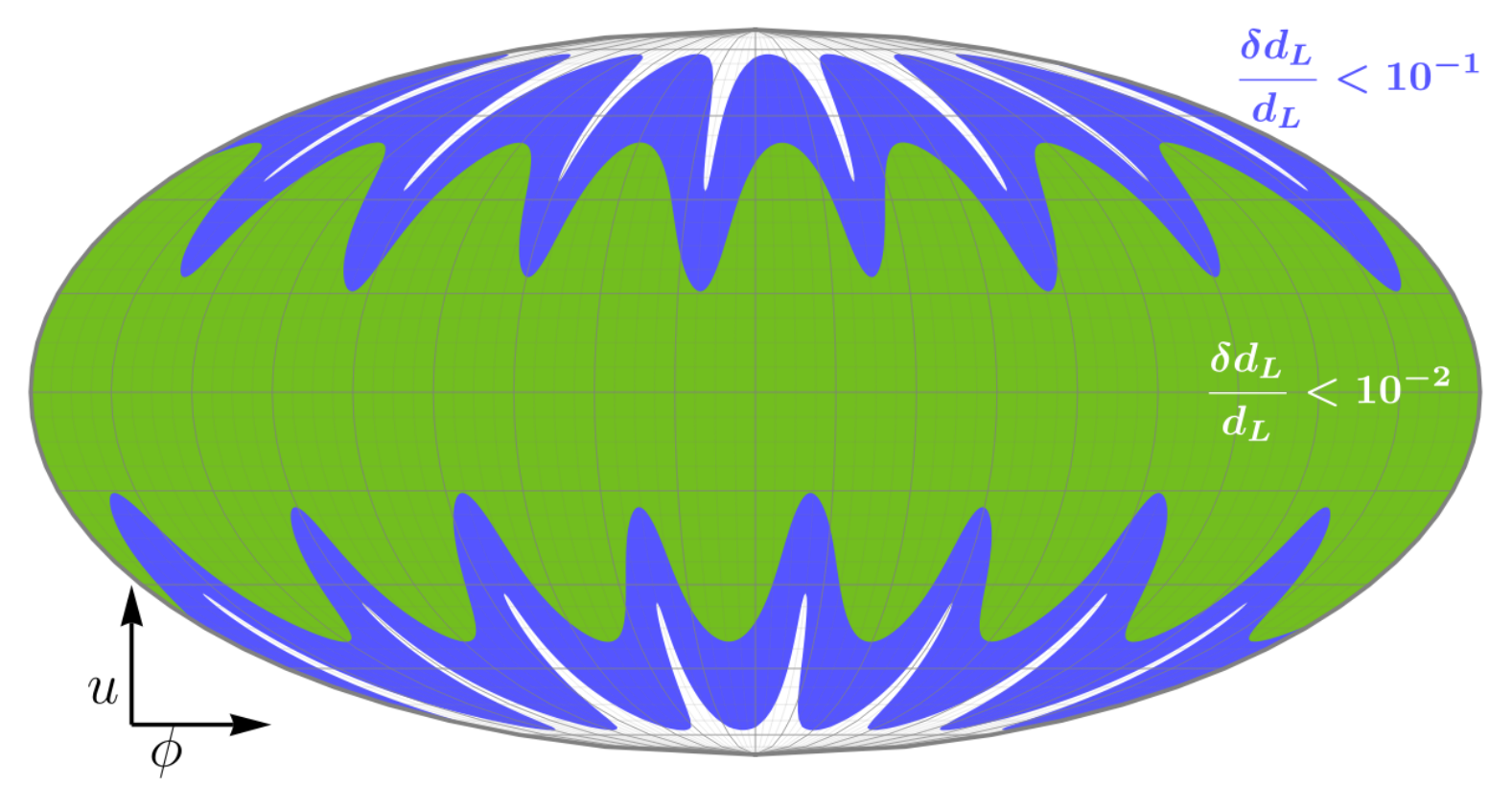}
\end{tabular}
  \caption{Dependence of the sky localization errors (top) and distance errors (bottom)  on sky position. Here we consider a binary with $M_s=10^7 M_\odot$ at $z=1$ with $(q,\,\iota,\,\psi)=(2,\,45\degree,\,60\degree)$. We also plot localization ellipses at constant $u$ and $\phi$. For visualization purposes we magnify the size of each ellipse by a factor $10$ (i.e., we magnify the area by a factor $100$).}
\label{fig:skyDep}
\end{figure}

\subsection{Sky-location dependence}

Figure~\ref{fig:skyDep} shows the dependence of the localization errors (top panel) and luminosity distance errors (bottom panel) for a remnant source mass $M_s=10^7 M_\odot$ with $z=1$ and $(q,\,\iota,\,\psi)=(2,\,45\degree,\,60\degree)$.
In this case the best sky localization (top panel) and distance determination (bottom panel) are achieved when the binary is near the equator.

This is in contrast with errors on the remnant mass, remnant spin, mass ratio and inclination, which are smaller when the source is overhead. The reason is that sky localization and distance determination hinge on measuring the relative amplitudes or phases between two channels. For overhead binaries the SNR is close to optimal, but both channels have similar amplitudes and phases. Consequently, localization is much better when the binary is close to the equatorial plane, even though the SNR is not optimal.

\section{Conclusions}
\label{sec:conclusions}

Massive BH binaries in the universe are expected to have a stronger influence on their astrophysical environment. Partly because of observational bias, there is by now strong observational evidence for BHs in the high-mass range, and mounting evidence that they may form binaries. For example, the Catalina Real-time Transient Survey (CRTS) identified $111$ candidate SMBH binaries with periodic variability~\cite{Graham:2015tba}, more than $90\%$ of which have masses $\gtrsim 10^8 M_\odot$. If even a small fraction of the high-mass BH binaries in the universe merge, higher modes of the ringdown may be detectable by LISA.

The ability to localize high-mass BH binaries is particularly important. If binary BH mergers are accompanied by electromagnetic signatures (like a ``notch'' in the IR/optical/UV spectrum, or periodically modulated hard X-rays), such signatures are most likely in massive binaries, with typical masses in the range $10^8 M_\odot$--$10^9 M_\odot$ (see e.g.~\cite{Krolik:2019nnw}). In particular, {\em Athena} should be able to detect X-ray emission from such sources at $z\lesssim 2$~\cite{AthenaLISA,McGee:2018qwb}. %
The coincident detection of gravitational and electromagnetic waves may allow us to use BH binaries as standard sirens at relatively large redshift~\cite{Holz:2005df,Tamanini:2016zlh}, potentially resolving the apparent discrepancy between cosmological observations at early and late cosmological time~\cite{Verde:2019ivm}.

In this paper we have shown that higher modes of the merger and ringdown are a treasure trove of information on various properties of the binary, such as the mass ratio, inclination, sky location and luminosity distance. This is particularly remarkable because the source localization method we proposed here (while admittedly somewhat limited in scope) {\em does not rely on modulations induced by LISA's motion}, and therefore it is independent of the observation time.

For the reader's convenience, we conclude this paper with a short summary of our main results.

In Sec.~\ref{sec:remnant} we use Fisher matrix estimates for the remnant mass and spin from past work [Eq.~(\ref{eq:mass-spin}): see e.g.~\cite{Echeverria:1989hg,Finn:1992wt,Berti:2005ys})], showing that the accuracy with which these parameters can be measured improves by combining several modes.

In Sec.~\ref{sec:qiota} we present one of our central results: since we know how the ringdown amplitudes depend on mass ratio, we can obtain {\em both the mass ratio and the inclination} of the binary from the measurement of three modes. The key insight comes from Eq.~(\ref{eq:Omegall}), which implies that by taking appropriate ratios of the three dominant modes we can find both $q$ and $\iota$.

In Sec.~\ref{sec:skyloc} we assume that $\iota$ has been determined as described in Sec.~\ref{sec:qiota}, and we show that multi-mode detections allow us to determine the sky localization and luminosity distance without having to rely on modulations induced by LISA's orbital motion. We define the ratio between the signal amplitudes in two LISA channels of detectors $Q^{\ell m}_A$ [Eq.~(\ref{eq:QAdef})] and the difference between their phases $Q^{\ell m}_\Phi$ [Eq.~(\ref{eq:QPdef})]. The two $\ell=m=2$ quantities $Q_A\equiv Q^{22}_A$ and $Q_\Phi\equiv Q^{22}_\Phi$ should typically be measured with the highest SNR, and they depend on three angles: $(\theta,\,\phi,\,\psi)$. For constant values of $Q_A$ and $Q_\Phi$, we can eliminate $\psi$ %
 and identify contours in the sky (Fig.~\ref{fig:QAP}).
A similar procedure can be applied to the relative channel power $Q_C$ [Eq.~(\ref{eq:Q1def})] and the relative polarization power $Q_P$ [Eq.~(\ref{eq:Q2def})], leading to the identification of additional ``rings in the sky'' (Fig.~\ref{fig:localizationRingsQC}). Finally, the intersection of these two sets of ``rings in the sky'' identifies finite sets of {\em points} where the source may be located. A similar strategy allows us to determine the luminosity distance (Sec.~\ref{sec:lumdistance}).
In Sec.~\ref{sec:qIotaSkyDep} we discuss how parameter estimation accuracy depends on the binary's mass ratio, inclination and sky position.

Our analysis relies on several simplifying assumptions that should be relaxed in future work. For example, we neglect the effect of spins on the mode amplitudes, which is reasonably well understood (see e.g.~\cite{Baibhav:2018rfk} and references therein). Spins should not significantly affect the errors on mass ratio $q$ and inclination $\iota$: these quantities depend on the amplitude ratios of $\ell=m$ modes, which are only mildly dependent on spins, as first shown by~\cite{Kamaretsos:2011um,Kamaretsos:2012bs}. The situation is different for the $(2,1)$ mode (crucial to estimate sky localization and luminosity distance), which is very sensitive to spins. In this case, correlations between the spins and other binary parameters could reduce the accuracy in sky localization and luminosity distance. However, by focusing on the ringdown we have significantly {\em underestimated} the information carried by the full inspiral-merger-ringdown signal, which should break some of these correlations. For example, LIGO observations of the inspiral can most easily measure the ``effective spin'' combination $\chi_{\rm eff}=(q\chi_1+\chi_2)/(q+1)$~\cite{Racine:2008qv,Kesden:2010yp}, while the $(2,\,1)$ mode depends most sensitively on the combination  $\chi_-=(q\chi_1-\chi_2)/(q+1)$~\cite{Baibhav:2017jhs}. Combined measurement of the inspiral and of the ringdown could reduce the errors on the individual spin components. These qualitative arguments should be supported by explicit calculations using state-of-the-art inspiral/merger/ringdown models including higher harmonics~\cite{OShaughnessy:2017tak,Kumar:2018hml,Cotesta:2018fcv,Mehta:2019wxm,Breschi:2019wki,Shaik:2019dym}, a task beyond the scope of this work.

\noindent{\bf{\em Acknowledgments.}}
We thank John Baker, Julian Krolik, Maria Okounkova and Alberto Sesana for discussions, and Monica Colpi for carefully reading an early version of this draft.
E.B. and V.B. are supported by NSF Grants No. PHY-1912550 and AST-1841358, NASA ATP Grants No. 17-ATP17-0225 and 19-ATP19-0051, and NSF-XSEDE Grant No. PHY-090003. E.B. acknowledges support from the Amaldi Research Center funded by the MIUR program ``Dipartimento di Eccellenza''~(CUP: B81I18001170001). 
V.C. acknowledges financial support provided under the European Union's H2020 ERC 
Consolidator Grant ``Matter and strong-field gravity: New frontiers in Einstein's 
theory'' grant agreement no. MaGRaTh--646597.
This work has received funding from the European Union’s Horizon 2020 research and innovation programme under the Marie Skłodowska-Curie grant agreement No. 690904. This research project was conducted using computational resources at the Maryland Advanced Research Computing Center (MARCC). The authors would like to acknowledge networking support by the GWverse COST Action CA16104, ``Black holes, gravitational waves and fundamental physics.''

\appendix

\section{Localization from time evolution of antenna pattern}
\label{sec:evolutionLocalization}

Most long-lived sources can be localized using the time variation of the LISA antenna pattern. This method cannot be used for ringdown waveforms, because they are short-lived: a typical ringdown decay time ranges from $1$ minute for $M_s\sim 10^6 M_\odot$ to $\approx 17$ hours for $M_s\sim 10^9 M_\odot$.  This is a problem for very massive BH mergers, where the inspiral occurs out of band and we may have to rely only on merger-ringdown to localize the source. %

Let us assume that the source direction remains constant in the Solar System frame during the observation period.
In the LISA frame, the position ${\bf r}$ of a GW source which has fixed position ${\bf r_B}$ in the Solar System is given by
${\bf r} = {\mathcal{R}}\cdot {\bf r_B}$,
where ${\mathcal{R}}(t)  =  \mathcal{C}\cdot\mathcal{B}\cdot\mathcal{A}$ is a product of three rotation matrices:
\bea
\mathcal{A} & = & \left(\begin{array}{ccc}
\cos\omega t & \sin\omega t & 0\\
-\sin\omega t & \cos\omega t & 0\\
0 & 0 & 1\end{array}\right) \, , \nonumber \\
\mathcal{B} & = & \left(\begin{array}{ccc}
1 & 0 & 0\\
0 & \frac{1}{2} & \frac{\sqrt{3}}{2}\\
0 & -\frac{\sqrt{3}}{2} & \frac{1}{2}\end{array}\right) \, , \nonumber \\
\mathcal{C} & = & \left(\begin{array}{ccc}
\cos\omega t & -\sin\omega t & 0\\
\sin\omega t & \cos\omega t & 0\\
0 & 0 & 1\end{array}\right) \, .
\eea
Here $\omega={2\pi}/{T}$ is the LISA orbital frequency, and $T=1$~yr. The source direction in the barycentric 
frame can be written in polar coordinates as $\hat{n}_B = (\sin\theta_{B}\cos\phi_{B}, \sin\theta_{B}\sin\phi_{B}, \cos\theta_{B})$, and the 
corresponding vector in the LISA frame is ${\bf \hat{n}_L}={\cal R}(t)\cdot{\bf \hat{n}_B}$.

In the LISA frame, the apparent change in position of the source is given by
\be
{\bf \hat{n}(t)}={\mathcal{R}}(t)\cdot{\mathcal{R}}(0)^{-1}\cdot \boldsymbol{\hat{n}(0)}\,.
\ee
If we apply this transformation to the source position vector we get
\bea
 \cos\theta(t) &=& \cos\theta +  \frac{\sqrt{3}}{2} \omega t \cos\phi \lvert\sin\theta\rvert\,,\nn\\
\phi(t) &=& \phi +\frac{1}{2}\omega t\left(1+\sqrt{3} \cot\theta \sin\phi\right)\,,
\eea
while if we apply it to the angular momentum vector of the binary 
${\bf \hat{ L}}_B = (\sin\theta_{L}\cos\phi_{L}, \sin\theta_{L}\sin\phi_{L}, \cos\theta_{L})$ we find that the inclination $\iota =\cos^{-1}({\bf \hat{L}_B}\cdot{\bf \hat{n}_B})$ is constant, while the polarization angle,
given in terms of ${\bf z}$ (the direction perpendicular to the LISA plane) by
\be
\tan \psi(t)= \f{-{\bf \hat{z}}\cdot \left({\bf \hat{n}}\times({\bf \hat{n}}\times{\bf \hat{L}})\right)}{{\bf \hat{z}}\cdot({\bf \hat{n}}\times{\bf \hat{L}})}\,,
\ee
changes at ${\mathcal O}((\omega t)^2)$. The waveform modes change as follows:
\bea
h_\lm(t) &=  \left[{\mathcal A}_\lm+ \omega t\,{\mathcal B}_\lm^1+ \f{1}{2}(\omega t)^2\,{\mathcal B}_\lm^2\right]e^{-(t-t_0)/\tau_\lm} \nonumber \\
&\times \cos\left [2\pi f_\lm t + \Phi_\lm + \omega t\,\Psi_\lm^1+ \f{1}{2}(\omega t)^2\,\Psi_\lm^2 \right]\,,\nonumber\\
\label{eq:hDefOrder1}
\eea
where the first-order corrections to the detector amplitude and phases are
\bea
{\mathcal B}_\lm^1 &=&\frac{{\mathcal A}_\lm}{\Omega_\lm} \f{1}{\omega} \f{d}{dt} \left( \sqrt{(F_+ (t)\, Y_+^\lm)^2 + (F_\times(t)\, Y_\times^\lm)^2}  \right)\,,\nn\\
 \Psi_\lm^1 &=& \f{1}{\omega} \f{d}{dt} \left( \f{F_\times(t)\, Y_\times^\lm}
{F_+(t)\, Y_+^\lm}  \right)\,,
\eea
and the second-order corrections are
\bea
{\mathcal B}_\lm^2 &=&\frac{{\mathcal A}_\lm}{\Omega_\lm} \f{1}{\omega^2} \f{d^2}{dt^2} \left( \sqrt{(F_+ (t)\, Y_+^\lm)^2 + (F_\times(t)\, Y_\times^\lm)^2}  \right)\,,\nn\\
 \Psi_\lm^2 &=& \f{1}{\omega^2} \f{d^2}{dt^2} \left( \f{F_\times(t)\, Y_\times^\lm}
   {F_+(t)\, Y_+^\lm}  \right)\,.
 \nn
\eea

By computing Fisher matrices, we can show that the first-order corrections can be measured with accuracy
\bea
\delta {\mathcal B}_\lm^1 &=& \f{\sqrt{2}}{\pi} \f{\mathcal{A}_\lm}{\rho_\lm}\f{T}{\tau_\lm}\,, \nn\\
\delta \Psi_\lm^1 &=& \f{\sqrt{2}}{\pi} \f{1}{\rho_\lm}\f{T}{\tau_\lm}\,,\nn\\
\delta {\mathcal B}_\lm^2 &=& \sqrt{\f{2}{3}}\f{1}{\pi^2}\f{{\mathcal A}_\lm}{ \rho_\lm}\left(\f{T}{\tau_\lm}\right)^2\,, \nn\\
 \delta \Psi_\lm^2 &=& \sqrt{\f{2}{3}}\f{1}{\pi^2}\f{1}{\rho_\lm}\left(\f{T}{\tau_\lm}\right)^2\,.
 \eea

\begin{figure}[t]
  \includegraphics[width=0.9\columnwidth]{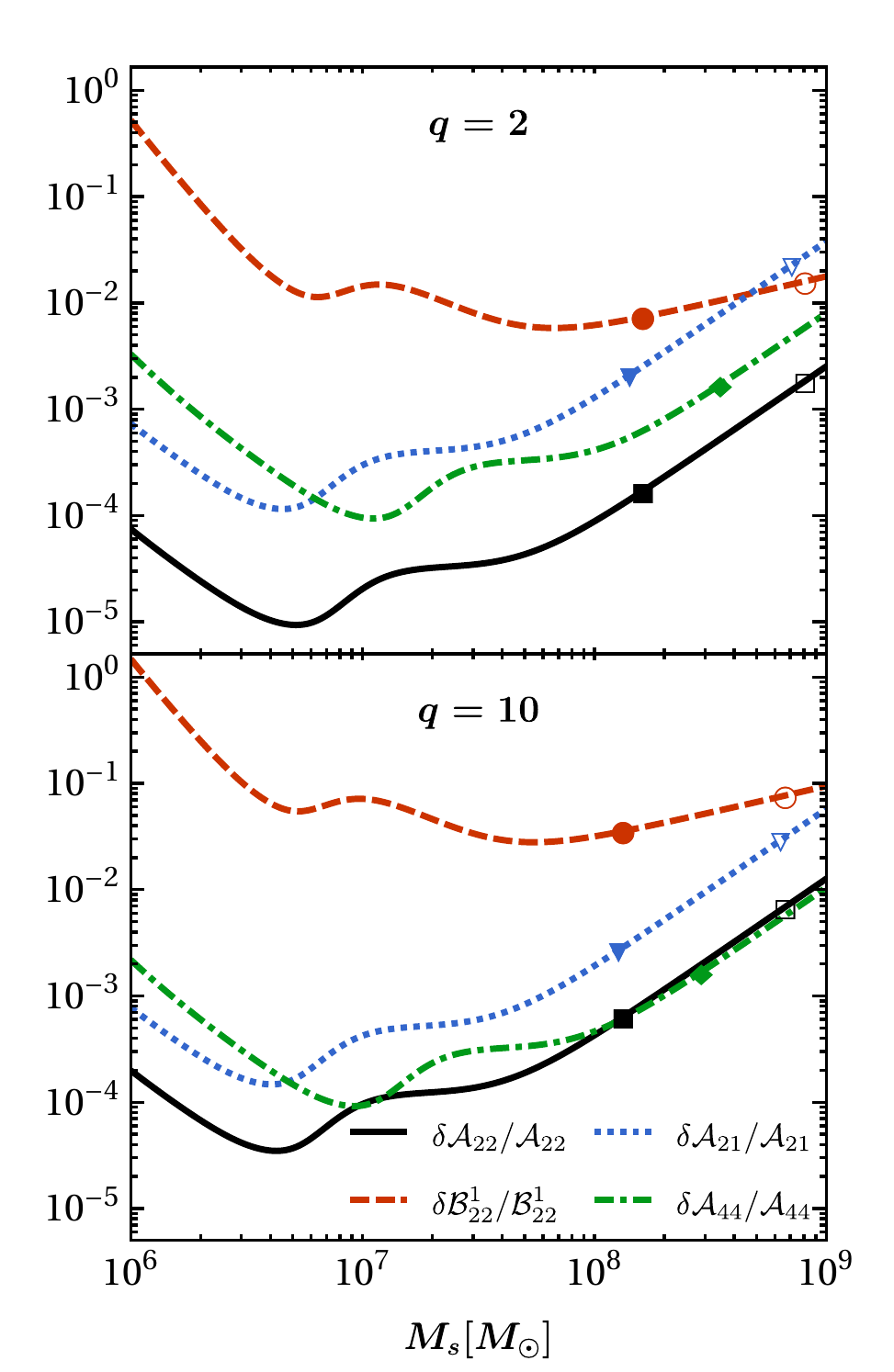}
  \caption{Fractional amplitude errors for a source at $z=0.1$. The markers indicate the mass at which the given mode goes out of band at $f_{\rm cut}=10^{-4}$~Hz (solid markers) and $f_{\rm cut}=2\times10^{-5}$~Hz (hollow markers).} 
\label{fig:errTs}
\end{figure}

For long-lived sources, the evolution of antenna pattern can be used to find both the inclination and the sky position. Recall however that our strategy in this paper relies on first using the $\ell=m=2,\,3,\,4$  modes to find the inclination, and then the $(2,\,1)$ mode to find the sky position. The question is then whether first-order in $\omega t$ corrections to the dominant mode amplitude $\delta {\mathcal B}_{22}^1$, which could be used to find the source position and orientation, can be measured more or less accurately than the other subdominant amplitudes $\mathcal{A}_{\ell m}$ themselves. In Fig.~\ref{fig:errTs} we plot the fractional error $\delta {\mathcal B}_{22}^1/{\mathcal B}_{22}^1$ and we compare it to $\delta \mathcal{A}_{\ell m}/\mathcal{A}_{\ell m}$ for some of the dominant $(\ell,\,m)$ modes for sources of different mass at redshift $z=0.1$.
For $q=10$ (bottom panel), $\delta {\mathcal B}_{22}^1/{\mathcal B}_{22}^1$  is larger than either $\delta \mathcal{A}_{44}/\mathcal{A}_{44}$ or $\delta \mathcal{A}_{21}/\mathcal{A}_{21}$ for all binaries with $M<10^9 M_\odot$, so the time evolution of the signal amplitude should not play an important role in finding inclination. Furthermore, in this paper,
for $q=2$ (top panel), $\delta \mathcal{A}_{21}/\mathcal{A}_{21}$ gets larger than  $\delta {\mathcal B}_{22}^1/{\mathcal B}_{22}^1$ when $M\gtrsim 5\times10^8 M_\odot$ and slight improvements in source localization may be possible. Note however that these improvements would only be possible if we can control the low-frequency sensitivity down to $f_{\rm cut}=2\times10^{-5}$~Hz. Solid markers in Fig.~\ref{fig:errTs} show that, if $f_{\rm cut}=10^{-4}$~Hz, the signal would get out of band before any improvement occurs. 

\section{Parameter estimation for sources with electromagnetic counterparts}
\label{sec:EM}
In this section we consider parameter estimation errors in the ideal situation where we can associate an optical counterpart to the source, so that $\theta,\,\phi$ and $d_L$ are known.

A single-mode detection is enough to solve for the remaining unknowns $(q,\,\iota,\,\psi)$. For example, from the knowledge of $(\theta,\,\phi)$ we can use $Q_A^\lm$ and $Q_\Phi^\lm$ to measure $\iota$ and $\psi$,
which can then be used to solve for $q$. We need a Jacobian transformation from the basis $\boldsymbol{Q_\lm}=\{Q_A^\lm,\, Q_\Phi^\lm\}$ to the basis  $\{\iota,\,\psi\}$, and we can propagate the uncertainty as usual:
\be
\cov(\{\iota,\psi\}_\lm)=\jacob{\{\iota,\psi\}}{\boldsymbol{Q_\lm}}\cdot \cov(\boldsymbol{Q_\lm})\cdot\left(\jacob{\{\iota,\psi\}}{\boldsymbol{Q_\lm}}\right)^T \,, 
\ee
where the covariance for $\boldsymbol{Q_\lm}$ is
\be
\cov(\boldsymbol{Q_\lm}) = \sum_{i={\rm I, II}} \f{1}{(\rho_\lm^i)^2} \begin{pmatrix} 
2 & 0 \\
0 &  1 
\end{pmatrix}\,.
\ee
The Jacobian for $\boldsymbol{Q_\lm}\to\{\iota,\,\psi\}$ is
\be
\jacob{\{\iota,\psi\}}{\boldsymbol{Q_\lm}}=\left(\jacob{\boldsymbol{Q_\lm}}{\{\iota,\psi\}}\right)^{-1}\,,
\ee
which can be calculated from Eq.~(\ref{eq:QAdef}) and Eq.~(\ref{eq:QPdef}).

We can then compute the reduced error as 
\be
\cov(\{\iota,\psi\}_\lm)=\left(\sum_\lm \left(\cov(\{\iota,\psi\}_\lm)\right)^{-1}\right)^{-1}\,.
\ee
Once $\iota$ and $\psi$ are known we can compute $q$ from
\be
A_\lm(q)=\f{{\mathcal A}_\lm d_L}{\Omega_\lm(\iota,\theta,\psi) M}\,,
\ee
and error propagation gives 
\be\label{eq:errqEM}
\delta q^2 =\left(\f{A_\lm(q)}{A_\lm'(q)}\right)^2\left[ \left(\f{\delta M}{M} \right)^2 +\f{2}{\rho_\lm^2 } +\left(\f{\delta \Omega_\lm}{\Omega_\lm}\right)^2 \right]\,,
\ee
where
\bea
\Omega_\lm^2&=& \left(\Omega^{\rm I}_\lm\right)^2 + \left(\Omega^{\rm II}_\lm\right)^2,\, \nn\\
\rho_\lm^2 &=& \left(\rho^{\rm I}_\lm\right)^2 + \left(\rho^{\rm II}_\lm\right)^2,\,
\eea
and
\be
(\delta\Omega_\lm)^2= \jacob{\Omega_\lm}{\{\iota,\psi\}}\cdot \cov(\{\iota,\psi\})\cdot\left(\jacob{\Omega_\lm}{\{\iota,\psi\}}\right)^T \,.
\ee

\begin{figure}[t]
\begin{tabular}{c}
  \includegraphics[width=0.95\columnwidth]{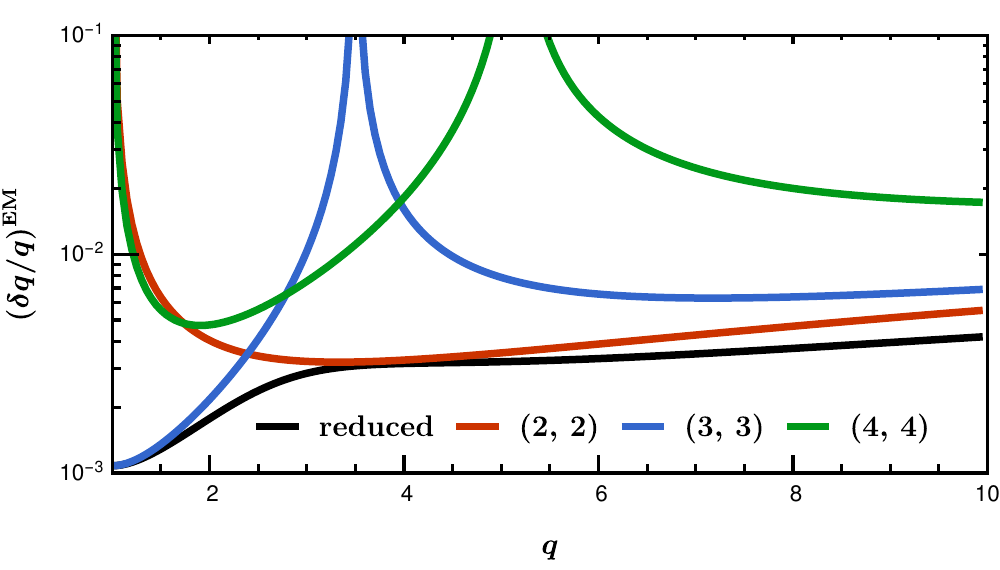}\\
  \includegraphics[width=0.95\columnwidth]{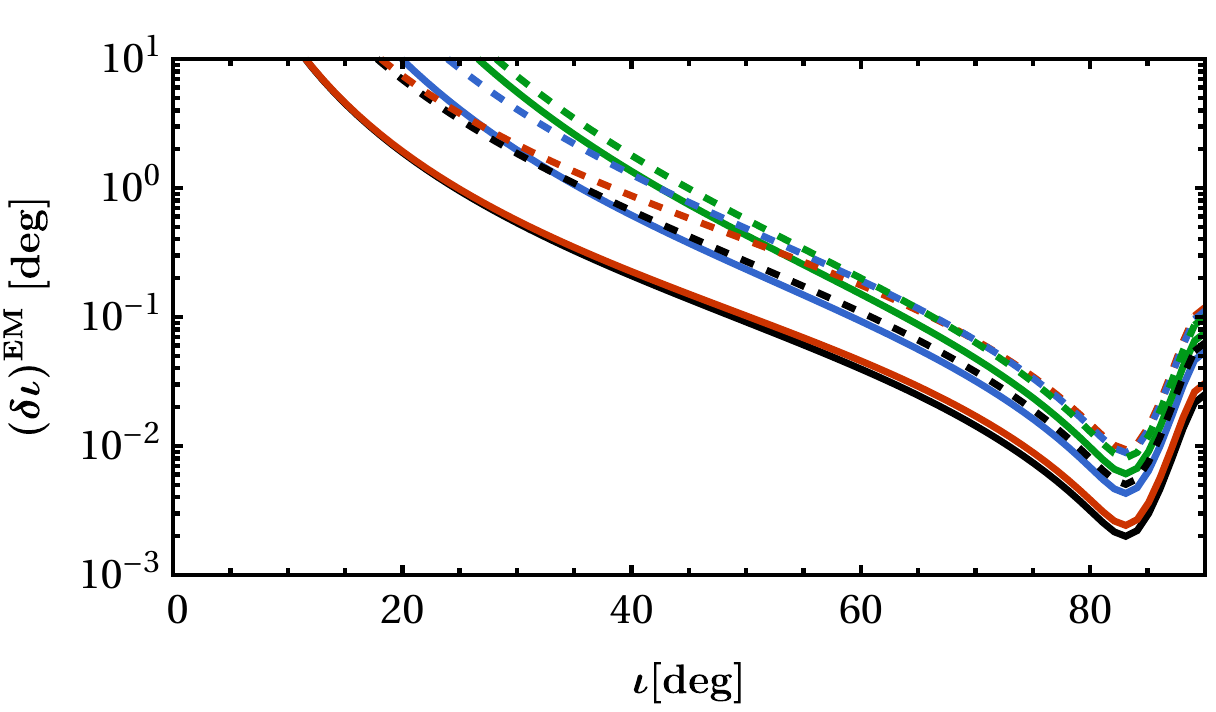}
  \end{tabular}
  \caption{Errors on $q$ and $\iota$ for a source with counterpart and $M_s=10^7 M_{\odot},\,z=1,\, u=0.5,\,\phi=30\degree,\,\psi=60\degree$. In the upper panel we set $\iota =45\degree$ while in the lower panel we set $q=2$ (solid lines) and $q=10$ (dashed lines).} 
\label{fig:QIdepEM}
\end{figure}

In Fig.~\ref{fig:QIdepEM} we plot the relative error on mass ratio
$\delta q/q$ and the inclination error $\delta \iota$ for a source
$M_s=10^7 M_{\odot}$ at $z=1$, assuming that the position and distance
of the source are known from an electromagnetic counterpart.

The upper panel of Fig.~\ref{fig:QIdepEM} shows that mass ratio errors
coming from a measurement of the $(2,\,2)$ and $(4,\,4)$ modes diverge
as $q\to 1$.  This is because $A_{22}'(q=1)=A_{44}'(q=1)=0$ and hence
the denominator in Eq.~(\ref{eq:errqEM}) diverges as $q\to1$. The
observed divergence of the errors for other modes and/or at other
values of $q$ are similarly due to the fact that
$A_\lm'(q)=0$. However, the solid black line shows that we can always
measure $q$ at the sub-percent level (at least in principle) by
combining information from all the modes.

The bottom panel of Fig.~\ref{fig:QIdepEM} shows that the inclination
is harder to measure for face-on binaries than for edge-on
binaries. This could be explained from a closer look at
Eq.~(\ref{eq:QAdef}) and Eq.~(\ref{eq:QPdef}). Note that $Q_A$ and
$Q_\Phi$ depend on inclination through the function $s_\lm$. As shown
in Fig.~\ref{fig:slm}, $s_\lm$ has a weak (strong) dependence on
$\iota$ for face-on (edge-on) binaries, leading to large (small)
errors. These considerations also apply to modes with $\ell>2$, which
in addition have smaller SNRs, and therefore larger errors. The
smaller SNR for edge-on binaries also leads to the observed turnover
for $\iota>80\degree$.

\bibliography{RingdownPE}
\end{document}